\documentclass[a4paper,fleqn,usenatbib]{mnras}

\usepackage{newtxtext,newtxmath}
\usepackage[T1]{fontenc}
\usepackage{ae,aecompl}

\usepackage{graphicx}	% Including figure files
\usepackage{amsmath}	% Advanced maths commands
\usepackage{amssymb}	% Extra maths symbols

\title[Planet-disc interactions in PDS~70]{Separating extended disc features from the protoplanet in PDS~70 using VLT/SINFONI}

\author[V.~Christiaens et al.]{V.~Christiaens$^{1,2,3}$\thanks{E-mail: Valentin.Christiaens@monash.edu}, S.~Casassus$^{1}$, O.~Absil$^{2}$\thanks{F.R.S.-FNRS Research Associate}, F.~Cantalloube$^{4}$, C.~Gomez Gonzalez$^{5}$, 
\newauthor{J.~Girard$^{6}$, R.~Ram\'irez$^{1}$, B.~Pairet$^{7}$, V.~Salinas$^{8}$, D.~J.~Price$^{3}$, C.~Pinte$^{3,5}$, S.~P.~Quanz$^{9}$,} 
\newauthor{A.~Jord\'an$^{10}$, D.~Mawet$^{11,12}$ and Z.~Wahhaj$^{13}$}
\\
$^{1}$Departamento de Astronom\'ia, Universidad de Chile, Casilla 36-D, Santiago, Chile\\
$^{2}$Space sciences, Technologies \& Astrophysics Research (STAR) Institute, Universit\'e de Li\`ege, All\'ee du Six Ao\^ut 19c, B-4000 Sart Tilman, Belgium\\
$^{3}$Monash Centre for Astrophysics (MoCA) and School of Physics and Astronomy, Monash University, Clayton Vic 3800, Australia\\
$^{4}$Max Planck Institute for Astronomy, Germany\\
$^{5}$Universit\'e Grenoble Alpes, IPAG, F-38000 Grenoble, France\\
$^{6}$Space Telescope Science Institute, 3700 San Martin Dr. Baltimore, MD 21218, USA\\
$^{7}$ISPGroup, ELEN/ICTEAM, UCLouvain, Belgium\\
$^{8}$Department of Physics and Astronomy, Graduate School of Science and Engineering, Kagoshima University, Japan\\
$^{9}$Eidgenossische Technische Hochschule Zurich Departement Physik, Institute for Particle Physics and Astrophysics, Zurich, Switzerland\\
$^{10}$Instituto de Astrof\'isica, Pontificia Universidad Cat\'olica de Chile, Vicu\~na Mackenna 4860, 7820436 Macul, Santiago, Chile\\
$^{11}$Department of Astronomy, California Institute of Technology, 1200 E. California Blvd, Pasadena, CA 91125, USA\\
$^{12}$Jet Propulsion Laboratory, 4800 Oak Grove Dr., Pasadena, CA 91109, USA\\
$^{13}$European Southern Observatory, Alonso de C\'ordova 3107, Vitacura, Santiago, Chile
}

\date{Accepted 2019 April 29. Received 2019 April 11; in original form 2019 February 20}

\pubyear{2018}

\begin{document}
\label{firstpage}
\pagerange{\pageref{firstpage}--\pageref{lastpage}}
\maketitle

\begin{abstract}
% 202/250 words
%The formation of planets is still mostly a theoretical field which requires critical input from observations. 
%Protoplanetary discs with inner clearings (transition discs) 
Transition discs are prime targets to look for protoplanets and study planet-disc interactions. 
We present VLT/SINFONI observations of PDS~70, a transition disc with a recently claimed embedded protoplanet. 
We take advantage of the angular and spectral diversity present 
in our data for an optimal PSF modeling and subtraction using principal component analysis (PCA).
We report the redetection of PDS~70~b, both the front and far side of the outer disc edge, and the detection of several extended features in the annular gap.
We compare spectral differential imaging applied before (PCA-SADI), and after (PCA-ASDI) angular differential imaging. 
Our tests suggest that PCA-SADI better recovers extended features, while PCA-ASDI is more sensitive to point sources.
We adapted the negative fake companion (NEGFC) technique to infer the astrometry of the companion, and derived 
$r = 193.5 \pm 4.9~\mathrm{mas}$ and PA $= 158.7\degr \pm 3.0$\degr.
We used both NEGFC and ANDROMEDA to infer the $K$-band spectro-photometry of the protoplanet,
and found results consistent with recent VLT/SPHERE observations, except for their 2018/02 epoch measurement in the $K2$ filter.
Finally, we derived an upper limit of $\dot{M_b} < 1.26 \times 10^{-7} \big[ \frac{5 M_{\rm Jup}}{M_b} \big] \big[ \frac{R_b}{R_{\rm Jup}}\big] M_{\rm Jup} $ yr$^{-1}$ for 
the accretion rate 
of the companion based on an adaptation of PCA-SADI/PCA-ASDI around the Br$\gamma$ line (assuming no extinction). % and adjacent channels. %, and the comparison of our non-detection with a recent H$\alpha$ detection.
%We extracted the spectrum of the candidate in the $K$ band and noticed that it was slightly brighter than the photometric points obtained by SPHERE/IRDIS in the $K1$ and $K2$ bands.
%We discuss possible reasons for this mild discrepancy.
%We cannot rule out the possibility that part of the signal from the location of the companion candidate stems from a filtered extended structure. 
%However, such gap-crossing filament would still physically require an embedded companion in the cavity. 
%We argue that {\bf the combination of our observations and recent images obtained with other high-contrast imagers} provide tentative hints supporting the companion+bridging material hypothesis.
\end{abstract}


%%%%%%%%%%%%%%%%%%%%%%%%%%%%%%%%%%%%%%%%%%%%%%%%%%

%%% abrreviated citations
\defcitealias{Muller2018}{M18}
\defcitealias{Keppler2018}{K18}

%%%%%%%%%%%%%%%%% BODY OF PAPER %%%%%%%%%%%%%%%%%%

\section{Introduction}

%Protoplanetary discs are the birth place of planets.
\emph{Transition discs} (TDs) are protoplanetary discs showing evidence of inner clearing. 
These inner clearings were first inferred from the SED, showing a lack of near- to mid-IR excess %\citep[e.g.][]{Strom1989,Skrutskie1990}
\citep[e.g.][]{Strom1989}, and recently confirmed by resolved sub-mm observations %\citep[e.g.][]{Andrews2011,Espaillat2014}. 
\citep[e.g.][]{Andrews2011}. 
\citet{Owen2016} recently highlighted the distinction between TDs with large gaps/cavities ($\gtrsim 20$ au), which show both high accretion rates and bright sub-mm fluxes, and mm-faint TDs which harbour smaller gaps.
TDs with large gaps/cavities appear incompatible with a photo-evaporated inner disc, with the cavity instead likely carved by embedded companions.
Yet, finding direct evidence of forming planets has proved difficult in these discs despite multiple claims \citep[][]{Kraus2012a,Quanz2013,Reggiani2014,Biller2014,Currie2015,Sallum2015,Reggiani2018}.
%Indeed, recent literature has shown that %the effect of 
The problem is that
aggressive filtering on bright and extended disc emission can create point-like artefacts, which could be confused with substellar companions, as 
seems to be the case in HD~169142 \citep[e.g.][]{Ligi2018}, in LkCa~15 \citep[][Currie et al. 2019, ApJL in press]{Thalmann2016}, and perhaps also in HD~100546
\citep{Follette2017, Rameau2017}.

While direct observational confirmation of giant planets in transition discs has turned out to be more difficult than expected, there is mounting indirect evidence of their presence \citep[e.g.][]{Casassus2016,Huang2018}. 
%In addition to the large gaps themselves,  and their possible discrepant morphologies at near-IR and sub-mm wavelengths \citep[e.g.][]{Garufi2013}, 
%further hints include asymmetric dust distributions \citep[e.g.][]{Casassus2013,van-der-Marel2013}, warps \citep[][]{Marino2015,Casassus2015} and spiral arms \citep[e.g.][]{Grady2001,Muto2012,Christiaens2014,Benisty2015}.
%While these features %(gaps, segregation of large and small grains, lopsided dust distributions and spiral arms) 
%can be produced through a range of different mechanisms, they also all have in common that they can be created by the presence of low-mass companions.
Large gaps of different sizes at near-IR and sub-mm wavelengths \citep[e.g.][]{Garufi2013}, 
asymmetric dust distributions \citep[e.g.][]{Casassus2013,van-der-Marel2013} %, warps \citep[][]{Marino2015,Casassus2015} 
and spiral arms \citep[e.g.][]{Muto2012,Christiaens2014,Benisty2015}
can be produced by a range of different mechanisms, but they are all potential by-products of companion-disc interactions. %the presence of companion(s) in the disc.
\citet{Price2018} recently showed that in the case of the transition disc around HD~142527, the characteristics of the disc including the gap size, the banana-shape mm-dust trap and the spiral arms at the gap edge, could all be qualitatively accounted for by the low-mass binary companion \citep[e.g.][]{Biller2012,Lacour2016}.
Could massive companions be at the origin of similar features observed in other transition discs? 
%Among the different potential signposts of planet presence, %to be found in the disc, 
%spiral arms might %be the features that could indirectly reveal 
%provide the most information about the companion that is launching them -- if that is indeed their origin.
%The global shape of the spiral \citep[][]{Ogilvie2002,Rafikov2002}, its pitch angle \citep[e.g.][]{Zhu2015a,Dong2015a}, its contrast \citep[][]{Dong2017a}, the number of observed arms \citep[e.g.][]{Dong2015a,Bae2017b} and the separation angle between primary and secondary arms \citep[][]{Fung2015,Bae2017b} can all be used to constrain the mass and location of the perturber. 
%Their global shape, pitch angle and contrast, the number of observed arms and the separation angle between primary and secondary arms can all be used to constrain the mass and location of the perturber \citep[e.g.][]{Ogilvie2002,Rafikov2002,Zhu2015a,Fung2015,Dong2017a,Bae2017b}.
%\citep[e.g.][]{Ogilvie2002,Rafikov2002,Zhu2015a,Fung2015,Dong2015a,Dong2017a,Bae2017b}. % \citep{Ogilvie2002,Rafikov2002,Zhu2015a,Dong2015a,Dong2017a,Fung2015,Bae2017b}. 

%{\bf ***Add paragraph on CPDs? ***}

High-contrast imaging is one of the most suitable techniques to detect faint signals such as newborn planets or circumstellar disc features in the close vicinity of young stars \citep[][]{Absil2010,Bowler2016}.
Differential imaging techniques allow to optimize the contrast reached at small angle. % \citep[see e.g.][]{Mawet2012b}.
Here, we focus on the combination of angular %differential imaging 
\citep[ADI;][]{Marois2006} and spectral differential imaging \citep[SDI\footnote{It was first referred to as \emph{spectral deconvolution} (SD) in those works to avoid confusion with dual-band (simultaneous) spectral differential imaging.};][]{SparksFord2002,Thatte2007}
%ADI exploits the rotation of the field of view occuring throughout an observation performed with the telescope pupil maintained fixed. 
%SDI exploits the radial motion of the stellar halo with wavelength, while the location of physical objects remains fixed. 
%This angular or spectral diversity enables the construction of a stellar halo model (subsequently subtracted to the images) containing as few physical signal from the vicinity of the star as possible.
%This model image can be built with algorithms involving LOCI \citep[][]{Lafreniere2007} or principal component analysis \citep[PCA;][]{Amara2012,Soummer2012}.
%The combination of ADI and SDI has been applied to adjacent spectral channels \citep[e.g.][]{Close2014} and low-spectral resolution integral field spectrographs (IFS) fed by extreme-AO instruments \citep[e.g.~][]{MacIntosh2015,Zurlo2016}.
%and report the first application of ADI+SDI on 
applied to medium-spectral resolution integral field spectrograph (IFS) data obtained with VLT/SINFONI on the transition disc of PDS~70. 

PDS~70 is a TTauri star surrounded by a disc with a remarkably large dust-depleted inner region. 
Previous literature has used a distance of $\sim$140~pc, assuming membership of PDS~70 to the Upper Centaurus Lupus association \citep{Riaud2006}.
However, the new Gaia DR2 distance places it at only $113\pm1$~pc \citep{Gaia2018}, which alter previous stellar parameter estimates.
The near-IR JHK spectrum of the star is compatible with an M1V pre-main sequence star \citep{Long2018}, although UBVRI photometry suggests a K5V star \citep{GregorioHetem2002}.
%An M1V spectral type would be compatible with a mass and age of $\sim$0.5 $M_{\odot}$ and $\sim 2$ Myr resp., 
%while a K5V spectral type would imply an older and more massive star \citep[$\sim$10~Myr and $\sim$1.1$M_{\odot}$ resp.;][]{Long2018}.
Based on a Markov-Chain Monte-Carlo method, \cite{Muller2018} provided new age and mass estimates of $5.4 \pm 1.0$ Myr and of $0.76 \pm 0.02 M_{\odot}$, resp., for the central star.
Considering the new distance throughout this paper, the size of the $\mu$m-size and mm-size dust cavities are $\sim$53 au and $\sim$65 au \citep{Hashimoto2012,Hashimoto2015,Long2018}.
Near-IR polarimetric images suggest an outer disc inclination of 45-50\degr~and a PA of the semi-major axis of $\sim$159\degr \citep{Hashimoto2012}.
SED modeling hints at the presence of a small optically thick inner disc \citep{Dong2012}.
The different gap sizes for $\mu$m-size and mm-size dust are suggestive of the presence of one or more companions inside the large annular gap \citep[e.g.][]{Pinilla2012,deJuanOvelar2013}.
A point source has been recently detected in the annular gap using VLT/SPHERE data, and interpreted as a forming planet \citep[][hereafter \citetalias{Keppler2018} and \citetalias{Muller2018}, resp.]{Keppler2018,Muller2018}.
The point source was tentatively re-detected in the H$\alpha$ line, suggesting on-going accretion and further supporting the protoplanet hypothesis \citep{Wagner2018a}.
While this work was under review, \citet{Keppler2019} presented new ALMA observations showing new substructures in the disc, some of which matched the NIR observations of \citetalias{Muller2018}.% and the tentative kinematic signature of a second companion at PA$
%sim$260\degr and 0\farcs39 projected separation.

In this paper, we combine spectral and angular differential imaging to reach high contrast in a VLT/SINFONI dataset acquired on PDS~70.
We first present our observations and data calibration procedure (Section~\ref{Obs+DataRed}).
We then explain our post-processing algorithms and show the resulting final images (Section~\ref{Postprocessing}).
%The optimal combination of reduction parameters that enables to maximize speckle subtraction and achieve the highest contrast in our post-processed images is detailed in Section~\ref{Algorithm}.
%Our final images are shown in Section~\ref{FinalImages}. 
We report the redetection of a point source at the location of the protoplanet, the redetection of both the front and far side of the outer disc edge, and the detection of several extended features in the annular gap, some of which present similarities to those claimed in \citetalias{Muller2018}. 
%During the finalization of the paper, ***mention SPHERE paper here if it comes out.
%The edge of the outer disc is also conspicuously re-detected, as in the first resolved IR images of the disc presented in \citet{Hashimoto2012}.
%Sec.~2 details the observations and data reduction.
%Our final images are presented in Sec.~3, and a brief analysis of the companion candidate and tentative spiral arm are provided
%Finally, our results are discussed and summarized in Sec.~4.
In order to better interpret our results, we injected both synthetic spiral arms and fake companions into our dataset and re-processed it with each of our post-processing algorithms. %(sections~\ref{SpiralInjectionTests} and \ref{CompanionInjectionTests}). 
%***Forward modeling?***
%To further help interpreting our results, we compared the spectrum retrieved at two locations in our images: at the protoplanet candidate location and along the bright edge of the outer disc (Section~\ref{NEGFCresults}).
Section~\ref{CompanionCandidate} is devoted to the characterization of the protoplanet in terms of astrometry, broad-band photometry, spectro-photometry and Br$\gamma$ emission.
%We then inferred the astrometry of the companion and its contrast with respect to the star as a function of wavelength (Section~\ref{NEGFCresults}).
%Finally, we also adapted our algorithm to look specifically for Br$_{\gamma}$ signatures (Section~\ref{BrgammaConstraints}).
Finally, we discuss the disc features identified in our images (Section~\ref{DiscFeatures}). %the potential of our two main post-processing algorithms, the features detected in our images and set constraints on the accretion rate of the protoplanet based on our non-detection (Section~\ref{Discussion}).
Our main conclusions are summarized in Section~\ref{Summary}.

\section{Observations and data reduction}\label{Obs+DataRed}

We observed PDS~70 with the AO-fed IFS VLT/SINFONI \citep{Eisenhauer2003,Bonnet2004} on May 10th 2014, as part of program 093.C-0526 (PI: S.~Casassus). 
A total of 116 datacubes (NEXP) were obtained, consisting of single exposures (NDIT=1) of 60s (DIT) each, hence amounting to 116min integration.
We took advantage of the pupil-tracking mode, which resulted in a total parallactic angle variation of 99.8\degr.
Observations were acquired in clear conditions with an average seeing (0\farcs7-0\farcs9 at $\lambda=$500 nm). %, ranging between 1\farcs0~ and 1\farcs3. 
The airmass spanned 1.06--1.20. 
We used the $H$+$K$ grating (spectral resolution $\sim$1500) and the 12.5mas plate scale, %(along the x and y axes resp.), 
resulting in a 0\farcs8 x 0\farcs8~field of view in each datacube.
However, in order to artificially increase our field of view and better sample the large annular gap in the disc, we followed a four-point dithering pattern throughout the observation, placing the star close to a different corner of the detector in consecutive exposures.

Data reduction followed the same procedure as described in \citet{Christiaens2018} for the HD~142527 dataset obtained the same night with the same observing strategy.
We refer to that work for details on SINFONI data calibration. % of the data. 
The only difference is that we median-combined each set of 4 consecutive cubes of PDS~70 %with a non-zero median 
after placing the star at the center of larger frames, such that the final calibrated cube has dimensions of 2000 x 29 x 101 x 101 (wavelength, time, $y$ and $x$ axes, resp.). 
For the centering, the centroid of the star was found in each frame by fitting a 2D Moffat function implemented in the Vortex Imaging Pipeline\footnote{Available at: \url{https://github.com/vortex-exoplanet/VIP}.} \citep[VIP;][Gomez Gonzalez et al. 2019, in prep.]{GomezGonzalez2017}.

\section{Post-processing}\label{Postprocessing}

This section focuses on the PCA-based algorithm we implemented to take advantage of both the spectral and angular diversity in our SINFONI data to produce high-contrast final images of the environment of PDS 70.
We also used the ANDROMEDA algorithm, which leverages the angular diversity in each individual spectral channels. We defer the interested reader to \citet[][]{Cantalloube2015} for a detailed description of ANDROMEDA.

\subsection{Algorithm description}\label{Algorithm}

Calibrated frames obtained with SINFONI are affected by more speckles in the stellar halo than images obtained with extreme-AO instruments.
Nonetheless, the advantage of SINFONI is its large wavelength coverage in the $H$+$K$ mode ($\sim$ 1.45-2.45 $\mu$m), where the stellar diffraction pattern (including speckles) moves radially by an amount proportional to the wavelength in different spectral channels.
To determine the optimal post-processing method for our 4D cube, we tested several variations of the PCA algorithms implemented in VIP, which are themselves based on the original PCA-ADI \citep{Soummer2012,Amara2012} and sPCA algorithms \citep{Absil2013a}.

We first compared (1) PCA in a single step considering both the angular and spectral variation together, (2) PCA-SDI + PCA-ADI in two consecutive steps (hereafter \emph{PCA-SADI}), and (3) PCA-ADI + PCA-SDI in two consecutive steps (hereafter \emph{PCA-ASDI}).
For both PCA-SADI and PCA-ASDI, the PCA-SDI part was performed in full frame, while the PCA-ADI part was performed in concentric annuli of 2-FWHM width where, for each annulus, the PCA library only includes frames where any putative companion would rotate by at least 1~FWHM.
This frame selection enabled us to minimize self-subtraction of faint signals of interest \citep{Absil2013a}.
PCA-SADI and PCA-ASDI consistently recovered the well-characterized western (forward-scattered) edge of the outer disc of PDS 70, and for some reduction parameters (detailed in the next paragraph) redetected either the faint far side of the disc (PCA-SADI) or the protoplanet candidate claimed in \citetalias{Keppler2018} and \citetalias{Muller2018} (PCA-ASDI).
By contrast, PCA in a single step did not unambigously detected the bright forward-scattered edge of the disc.  
We thus discarded PCA in a single step for the rest of the analysis.
%We interpret the poorer results obtained with PCA in a single step as due to speckles being significantly more correlated spectrally than temporally. 
%Division in two steps takes better advantage of the different speckle correlation levels, while a single larger PCA library is diluting the high spectral cross-correlation which enables an efficient speckle subtraction in the PCA-SDI step. 
%PCA-SADI and PCA-ASDI appear thus to be the most appropriate methods for SINFONI datasets. 
%It is noteworthy that this conclusion might not apply for extreme-AO fed IFS such as those of SPHERE/IFS, GPI or SCExAO/CHARIS because 1) the better AO system provides more consistent high Strehl ratio images through time, and 2) the IFS configuration is different (lenslet-based vs image-slicer based for SINFONI) so that the spectral correlation of speckles might be different \citep[e.g.][]{Claudi2008,Wolff2014,Brandt2017}.

Next, we tested different minor variations for the PCA-SADI and PCA-ASDI algorithms. We specifically tested:
\begin{enumerate}
\item using $K$-band spectral channels only, which benefit from a significantly better AO correction than the $H$-band channels, instead of using all $H$+$K$ channels;
\item normalizing spectral channels based on the stellar flux measured in a 1FWHM-aperture prior to the PCA-SDI part of either PCA-SADI or PCA-ASDI; 
\item using a minimum radial motion for the PCA-SDI part to only include spectral channels that would limit self-subtraction \citep[see also][]{Thatte2007};
\item using a maximum threshold in azimuthal motion when building the PCA-ADI library in order to limit the inclusion of temporally decorrelated speckles;
\item collapsing residual frames (obtained after PCA modeling and subtraction) using either a simple median or a variance-based weighted average \citep{Bottom2017}, for both the PCA-ADI and PCA-SDI parts. For this purpose, we implemented the SDI equivalent of the algorithm presented for ADI residuals in \citet[][]{Bottom2017}.
\end{enumerate}
We determined the optimal combination of reduction parameters based on the recovery of known key features of the disc in our post-processed image: the bright forward-scattered edge of the outer disc (West), the dim back-scattered edge of the outer disc (East), and a bright blob $\sim$0\farcs19 to the SE of the star \citep[][\citetalias{Keppler2018,Muller2018}]{Hashimoto2012}. The conclusions from our tests are as follows: 
\begin{enumerate}
\item No PCA-SADI and PCA-ASDI reduction using all $H$+$K$ channels managed to recover the faint back-scattered edge of the outer disc, in contrast to PCA-SADI reductions using only $K$-band channels. This suggests that the speckle subtraction during PCA-SDI is optimal when the high spectral correlation of the better AO-corrected $K$-band channels is not diluted in a larger PCA library. The effect of a better speckle subtraction appears to dominate over both the enhanced self-subtraction (due to less radial motion in SDI) and the lower number of accumulated photons (due to the use of only half of all spectral channels).
\item Normalizing spectral channels in flux led to a better redetection of the far side of the disc in the case of PCA-SADI, but did not improve the redetection of the blob in PCA-ASDI images. Since we will infer the spectro-photometry of the companion in Section~\ref{Spectro-photometry} using an adaptation of PCA-ASDI, we only normalize spectral channels in the case of PCA-SADI.
\item Using a minimum radial motion for the creation of the PCA-SDI library on $H$+$K$ channels implies the inclusion of comparatively more $H$-band spectral channels in the PCA library of $K$-band channels, and vice versa. %For our SINFONI data, this 
This leads to a worse speckle subtraction due to worse AO performance in $H$ band than in $K$ band - and hence corresponding decorrelation of speckles. 
Limiting our analysis to $K$-band channels, we built our PCA-SDI library using only the shortest wavelength channels of that band which kept a high speckle correlation with all other $K$-band channels and did not suffer from significant telluric absorption (spectral channels from 1.93 to 1.95 $\mu$m). The choice of these channels is further justified in Section~\ref{NEGFCresults}. %As shown in Figure~***, this choice will be further justified by the absence of signal above the speckle noise at the location of the companion candidate in PCA-ADI images of those spectral channels (contrary to PCA-ADI images of channels longward of $\sim$2.0 $\mu$m).
\item Using a maximum threshold for azimuthal motion to build the PCA-ADI library generally yields a better speckle subtraction \citep[e.g.][]{Wagner2018a}. However, with only 29 temporal cubes spanning $\sim$100\degr~rotation, adding this constraint to the minimum azimuthal threshold significantly decreases the number of principal components that can be used for ADI (e.g.~$n_{\rm pc}^{\rm ADI} \lesssim 3$ for a 1--3 FWHM azimuthal motion range). We noticed that using 3--10 principal components without a maximum rotation constraint led to a more conspicuous redetection of the companion using PCA-ASDI (Section~\ref{FinalImages}).
\item We did not obtain a significant improvement in the quality of post-processed images using a variance-based weighted average of residuals compared to the median. However, we later noticed that the variance-based weighted average led to a more accurate recovery of spectra of injected fake companions than the median, using the technique presented in Section~\ref{NEGFCresults}. %By contrast, the bright forward-scattered edge of the outer disc appears slightly more patchy using the variance-based weighted average of both PCA-ADI and PCA-SDI residuals.
\end{enumerate}
Our final choice of post-processing parameters for PCA-SADI and PCA-ASDI is thus the following: (i) we only consider $K$-band spectral channels, (ii) we normalize spectral channels in flux before PCA-SADI but not for PCA-ASDI, (iii) we set manually the PCA reference library to spectral channels between 1.93 and 1.95 $\mu$m for the SDI part, (iv) we do not set a maximum azimuthal motion to select frames for the ADI part (but do set a minimum threshold equivalent to a 1-FWHM linear motion), and (v) we consider the variance-based weighted average of residuals (for both the ADI and SDI parts) of channels longward of 2.0$\mu$m, 

Finally, in addition to PCA-SADI and PCA-ASDI, we also tested PCA-ADI in concentric annuli on the two cubes obtained after collapsing the spectral channels of the $H$ and $K$ bands separately.
No significant signal was found in the PCA-ADI $H$-band image, likely due to the poorer AO correction. However, this reduction still enabled us to set an independent upper limit on the $H$ band flux of the companion (Section~\ref{NEGFCresults}).
In contrast, PCA-ADI on the collapsed $K$ band yielded a tentative redetection of both the forward-scattering edge of the outer disc and the companion. %, and was used for comparison to PCA-SADI and PCA-ASDI regarding the potential for recovery of extended features and fake companions (section).

%\section{Results}

\begin{figure*}
	\includegraphics[width=0.99\textwidth]{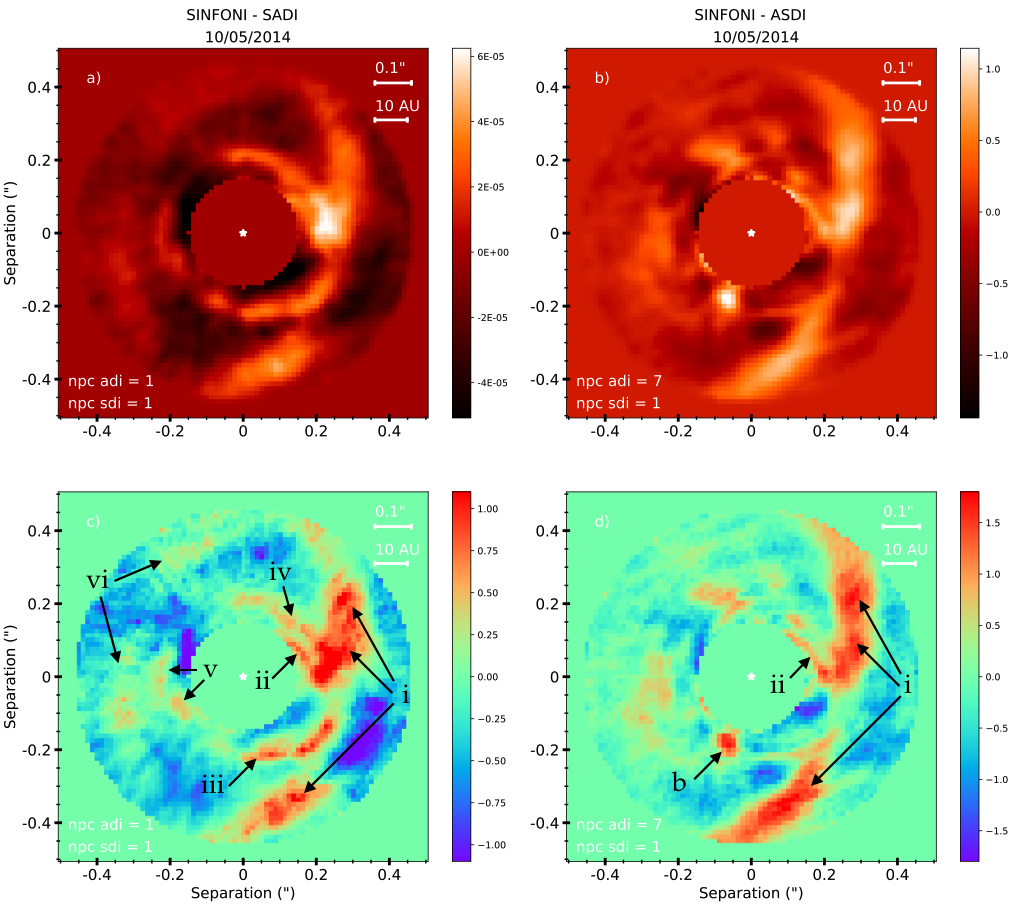}
%    \caption{Final images obtained with {\bf (a)} PCA-ASDI on the full $H$+$K$ 4D datacube, {\bf (b)} PCA-ADBI between the $K$- and $H$ bands, and {\bf (c)} PCA-ADI in concentric annuli in the $K$ band. See Sec.~\ref{Obs+DataRed} for details on the algorithms used. {\bf (d), (e)} and {\bf (f)}: STIM maps of (a), (b) and (c) respectively, as defined in \citet{Pairet2018}. Significant (\emph{i}, \emph{ii}, \emph{iii} and \emph{b}) and tentative features (\emph{iv?}, \emph{v?}, \emph{vi?}) are labelled in the STIM maps and described in the text. %{\bf See text for details on the computation of the STIM maps.} %as defined in \citet{Mawet2014}, but using the final image ob. 
%    The scale of all images is linear.}
    \caption{Final images obtained with {\bf (a)} PCA-SADI and the minimum number of principal components for both the ADI and SDI parts ($n_{\rm pc}^{\rm ADI}  = n_{\rm pc}^{\rm SDI}  = 1$), %{\bf (b)} PCA-SADI with $n_{\rm pc}^{\rm SDI}  = 1$ and $n_{\rm pc}^{\rm ADI}  = 7$, 
    and {\bf (b)} PCA-ASDI with $n_{\rm pc}^{\rm ADI}  = 7$ and $n_{\rm pc}^{\rm SDI}  = 1$. See Section~\ref{Postprocessing} for details on the algorithms used. {\bf (c)} and {\bf (d)}: STIM maps of (a) and (b) resp., enabling to highlight significant features in our final images \citep{Pairet2018}. A numerical mask of 0\farcs16 is used in all panels. The color scale of all panels is linear, between minimum and maximum values. Values are smaller in panel a because spectral channels are normalized before PCA-SADI (Section~\ref{Algorithm}). For the STIM maps, the maximum value of the color scale is set to the maximum value obtained in the respective inverse STIM maps (Figure~\ref{Inv_STIM_maps}): 1.1 and 1.8 for PCA-SADI and PCA-ASDI, respectively. This choice enables to highlight significant (\emph{i}, \emph{ii}, \emph{iii} and \emph{b}) and tentative features (\emph{iv}, \emph{v}, \emph{vi}) in our images. %See Section~\ref{FinalImages} for a description of the features. %{\bf See text for details on the computation of the STIM maps.} %as defined in \citet{Mawet2014}, but using the final image ob. 
    }
    \label{PCA-ASDI_final_img}
\end{figure*}

\subsection{Final images}\label{FinalImages}

Figures~\ref{PCA-ASDI_final_img}a and \ref{PCA-ASDI_final_img}b show the images obtained with our final choice of post-processing parameters for PCA-SADI and PCA-ASDI (Section~\ref{Algorithm}), respectively.
In order to assess the significance of features in our images, we show standardized trajectory
intensity mean (STIM) maps \citep{Pairet2018} in the lower two panels of Figure~\ref{PCA-ASDI_final_img}, preferring these over classical SNR$_t$ maps \citep{Mawet2014}.
This choice is detailed in Appendix~\ref{SNR_vs_STIM}. 
STIM maps are 2D detection maps defined at each location ($i$,$j$) as $\mu(x_{i,j})/\sigma(x_{i,j})$, where $\mu(x_{i,j})$ and $\sigma (x_{i,j})$ are the mean and standard deviation of trajectory $x_{i,j}$ throughout the derotated cube of residual images (i.e.~the cube of images obtained after PCA modeling and subtraction, and subsequent alignment with North up and East left), respectively.
Using STIM maps overcomes the difficulties of SNR$_t$ maps to identify significant signals at short separation from the star and in presence of bright disc features spanning a range of radii (as they bias the noise level estimation at each radius).

The STIM maps corresponding to the final PCA-SADI and PCA-ASDI reductions are shown in figure \ref{PCA-ASDI_final_img}c and d, respectively. 
The maximum value of the STIM map color scales corresponds to the maximum pixel value obtained in the respective inverse STIM maps (Figure~\ref{Inv_STIM_maps}).
The inverse STIM map is the detection map obtained by derotating the residual images using opposite parallactic angles. 
Since this procedure does not sum constructively authentic circumstellar signal while preserving the time dependence of residual speckles, it enables us to estimate the maximum pixel value that can be obtained in the detection map purely as a result of combining residual speckle noise \citep{MaroisLafreniere2008,Wahhaj2013}. 
In other words, pixels with values above that threshold in our (correctly derotated) STIM maps are unlikely to trace spurious signals. 
The threshold values are 1.1 and 1.8 in the PCA-SADI and PCA-ASDI STIM maps, respectively.
In Figure~\ref{PCA-ASDI_final_img}c and d, we identify Features \emph{i}, \emph{ii} and \emph{iii} and \emph{b} as significant, i.e.~containing pixels above that threshold.
We identify additional tentative Features \emph{iv}, \emph{v}, and \emph{vi}, (i.e.~containing pixels slightly below that threshold) based on the presence of similar features in the long-integration $K$-band image obtained with SPHERE/IRDIS \citepalias{Muller2018}.

Feature~\emph{b} coincides with the protoplanet claimed in \citetalias{Keppler2018,Muller2018} and \citet[][]{Wagner2018a}. %, and is analyzed in more details in Section~\ref{CompanionCandidate}.
Feature~\emph{i} corresponds to the bright forward-scattered edge of the outer disc, which has been identified in previous observations of the disc \citep[e.g.][\citetalias{Keppler2018}]{Hashimoto2012,Long2018}.
Feature~\emph{ii} has been identified as a possible gap-crossing bridge in \citetalias{Muller2018} (a blend of Features (2) and (3) in their figure B1).
Feature~\emph{iii} appears to connect the outer disc and the protoplanet, but has no correspondence in the images of \citetalias{Muller2018}.
Features~\emph{iv} and \emph{v} appear to match signals in the vicinity of Features (3) and (5) in figure B1 of \citetalias{Muller2018}.
Finally, Feature~\emph{vi} is consistent with the dim back-scattered light from the far side of the outer disc, also reported in \citetalias{Muller2018}.
Feature~b is analyzed in details in Section~\ref{CompanionCandidate}, while extended features are further discussed in Section~\ref{Disc-ussion}.

%Evolution of images with number of principal components

\begin{figure}
	\centering
	\includegraphics[width=0.475\textwidth]{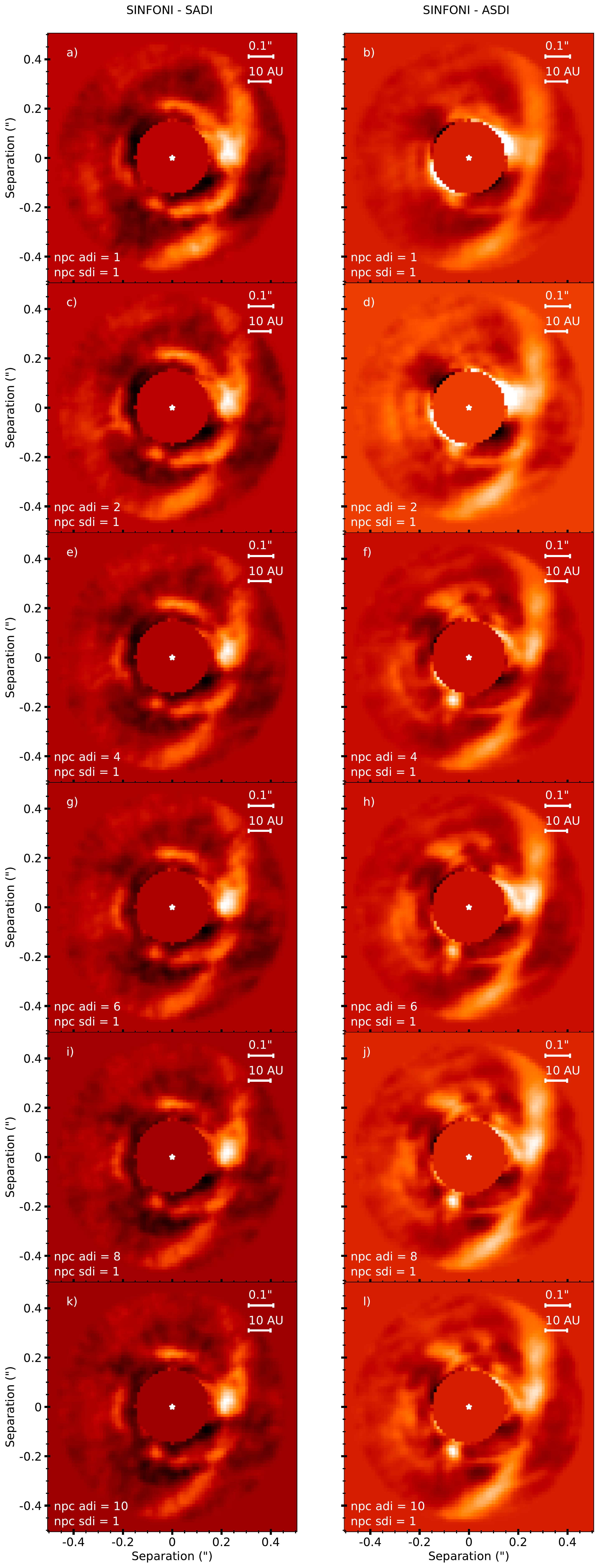}
    \caption{PCA-SADI (\emph{left}) and PCA-ASDI (\emph{right}) images obtained with different number of principal components for the PCA-ADI part ($n_{\rm pc}^{\rm ADI}$). All extended disc features are recovered by PCA-SADI for all values of $n_{\rm pc}^{\rm ADI} \in [1,10]$. The protoplanet is recovered by PCA-ASDI when $n_{\rm pc}^{\rm ADI} \ge 3$.}
    \label{npcADI_evo}
\end{figure}

In Figure~\ref{PCA-ASDI_final_img}, the choice of the number of principal components ($n_{\rm pc}$) is such that the detection of the faint back-scattered edge of the outer disc and the redetection of the protoplanet candidate are optimal in the PCA-SADI and PCA-ASDI images, respectively.
For comparison, images obtained with other values of $n_{\rm pc}^{\rm ADI} $ are shown in Figure~\ref{npcADI_evo}.
All the features identified in Figure 1c and d are also qualitatively recovered for all values of $n_{\rm pc}^{\rm ADI} $ from 1 to 10, albeit with increasing self-subtraction affecting extended features for increasing $n_{\rm pc}^{\rm ADI} $.
In particular, the faint far side of the disc can only be seen for the lowest $n_{\rm pc}^{\rm ADI} $ values for PCA-SADI, but is hardly seen in all PCA-ASDI reductions.
PCA-ASDI images with low $n_{\rm pc}^{\rm ADI}$ values also appear to leave stronger residuals at the edge of the 0\farcs16 numerical mask.
Only PCA-ASDI reductions with sufficiently large values of $n_{\rm pc}^{\rm ADI} $ ($\ge$ 3) enable to redetect the protoplanet candidate claimed in \citetalias{Keppler2018,Muller2018} and \citet{Wagner2018a}.
Nonetheless, these images also show some extended structures at $\sim$ 0\farcs2 to the NNE and at$\sim$ 0\farcs3 to the SE of the image, none of which being recovered in the PCA-SADI images. %, which suggest they might correspond to spurious artefacts.

Although not shown here, the results obtained with $n_{\rm pc}^{\rm SDI}  = 2$ are very similar to those obtained with $n_{\rm pc}^{\rm SDI}  = 1$. 
Increasing $n_{\rm pc}^{\rm SDI} $ to larger values leads to increasing self-subtraction.

Altogether, Figures~\ref{PCA-ASDI_final_img} and \ref{npcADI_evo} suggest that PCA-SADI with minimal number of principal components yields the best subtraction of speckles while minimizing self-subtraction of authentic signals, as it recovers better the faint far side of the disc.
Nonetheless, the fact that only PCA-ASDI with large $n_{\rm pc}^{\rm ADI} $ recovers a bright blob at the location of the previously claimed protoplanet candidate is puzzling.
Could Feature $b$ trace (at least partially) a co-located extended disc feature, possibly filtered into a point-like source by ADI processing? %\citep[as claimed e.g.~for HD 100546 b in][]{Rameau2017,Follette2017}?
To help answer this question, we performed two sets of tests; post-processing our data after injection of either extended or point-like features in our pre-processed data cube.
The results of these tests are presented in the next two sections.
%In the following section, we investigate the effect of each algorithm on fake injected extended signals. 

\begin{figure*}
	\centering
	\includegraphics[width=\textwidth]{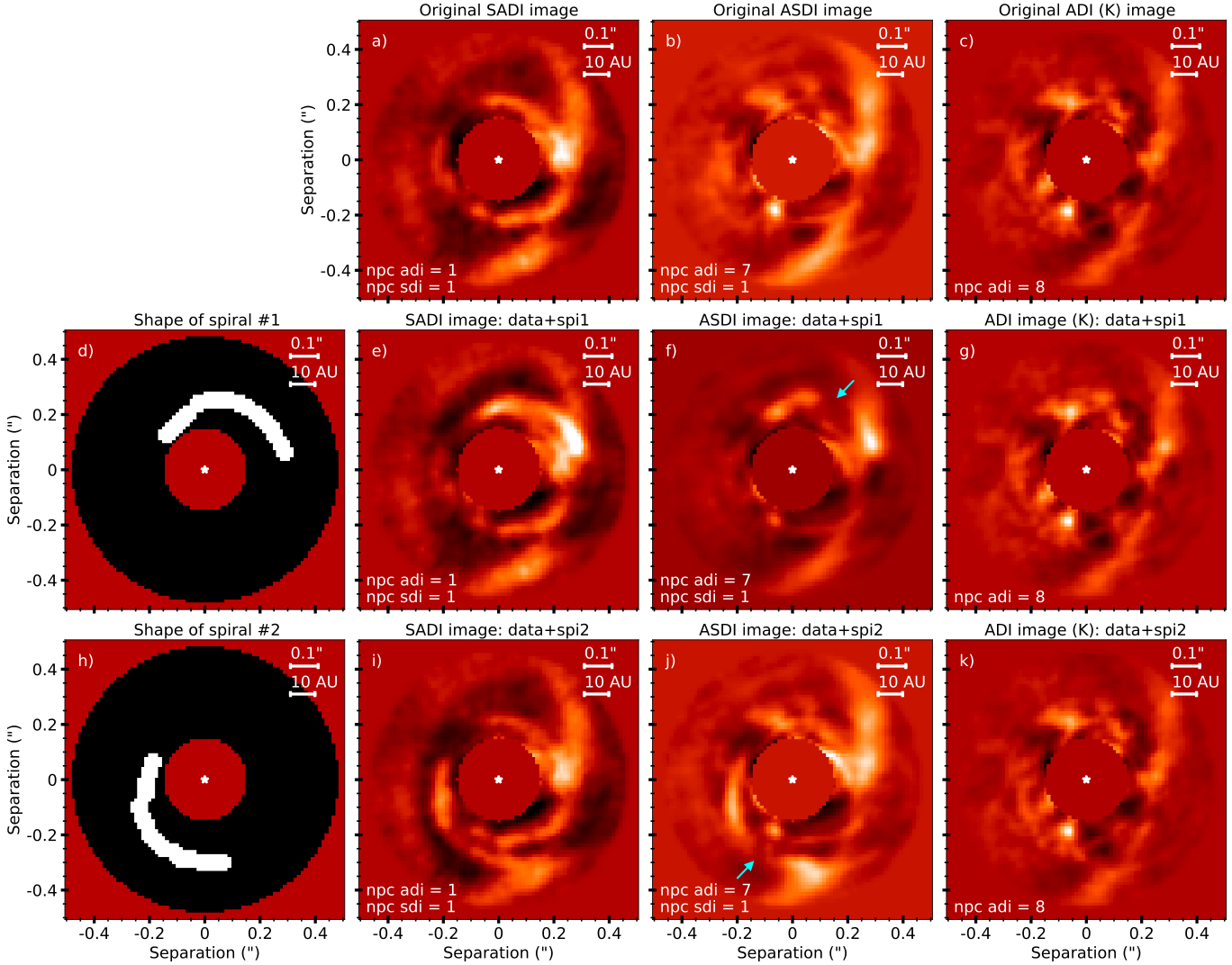}
    \caption{Tests of synthetic spiral arm injection. (\emph{Top row}) Final post-processed images obtained without the injection of any synthetic spiral, for the different algorithms tested in this paper: PCA-SADI, PCA-ASDI and PCA-ADI (left to right). (\emph{Middle} and \emph{bottom row}) Images obtained after using the same algorithms, but after injection of a synthetic spiral arm in the original 4D cube, with the shape given in the leftmost column (d and h).
%({\bf d}) and ({\bf h}): Shape of the synthetic spiral arms injected in the original 4D cube. ({\bf e}), ({\bf f}) and ({\bf g})  PCA-SADI, PCA-ASDI and PCA-ADI (K) images obtained after injection of synthetic spiral \#1 in the original 4D cube, respectively. ({\bf i}), ({\bf j}) and ({\bf k})  PCA-SADI, PCA-ASDI and PCA-ADI (K) images obtained after injection of synthetic spiral \#2 in the original 4D cube, respectively. %The spiral model used for injection is the best-fit spiral model shown in Figure~\ref{PolarImgs}a, but mirrored along axis PA = 275 \degr. It is fully recovered only in panel d. (\emph{Bottom row}) Final post-processed images obtained after injection of a synthetic spiral injected at the same level as the tentative observed spiral-like feature. Again, 
    The \emph{cyan arrows} show the locations where PCA-ASDI has significantly filtered out part of the injected spiral arm.
    The PCA-SADI reductions (e and i) recover better the full trace of the injected spiral. %Flux scales and limits are fixed in each column of images. 
    }
    \label{SpiralTests}
\end{figure*}

\subsection{Effect of post-processing on spiral arms} \label{SpiralInjectionTests}

% Following paragraph was originally in appendix
\subsubsection{Geometric biases}

%This section aims 
To evaluate how our PCA algorithms %the geometric biases introduced in our final images as a consequence of the algorithms we used, and specifically whether the latter could significantly 
filter out extended signals,
%To this purpose, w
we injected fake spiral arms in copies of our calibrated 4D cube and re-processed these cubes with PCA-SADI, PCA-ASDI and PCA-ADI (on the collapsed $K$-band spectral channels).
Results are shown in Figure~\ref{SpiralTests}.
%These tests aim to better evaluate possible geometric biases introduced with either PCA-SADI or PCA-ASDI. %reviously reported in \citet{Rameau2015}
We base the shape of the injected synthetic spirals on the best-fit of the trace of Feature~\emph{iii} (Figure~\ref{PCA-ASDI_final_img}c) to the linear spiral density wave equation, obtained when forcing the perturber location to be that of protoplanet candidate PDS 70 b \citep[e.g.][]{Rafikov2002,Muto2012}. %the observed tentative spiral arm seen between 0\farcs22 and 0\farcs31 separation to the S-SW in Figure~\ref{PCA-ASDI_final_img}a, but are injected mirroring the observed one with respect to axis PA = 275\degr (this model is best seen in Figure~\ref{SpiralTests}d).
The exact shape of the two injected synthetic spiral arms correspond to that spiral model flipped with respect to the x-axis (spiral \#1; Figure~\ref{SpiralTests}d), and rotated by 45\degr~clockwise (spiral \#2; Figure~\ref{SpiralTests}h), respectively.
%This enables to check in particular whether the effect of ADI %on the bright edge of the outer disc 
%can truncate the spiral.
The synthetic spirals are injected at roughly the same contrast level as Feature~\emph{iii}.
We considered a flat contrast at all wavelengths (i.e.~the spectrum of the injected spirals are a scaled down version of the stellar spectrum).
%We estimated the contrast of Feature~\emph{iii} to be $\sim$ 7.4 mag between PA $\sim 180$\degr and 225\degr, and injected our synthetic spirals at the same contrast level throughout the trace.
%We ran two sets of tests, one with a synthetic spiral that is $\sim$2 mag brighter than the observed one that enables to better visualize the effect of filtering (middle row of Figure~\ref{SpiralTests}), and a second set of tests where the synthetic spiral is injected at the same level as the observed one (bottom row of Figure~\ref{SpiralTests}).
For comparison, the first row of Figure~\ref{SpiralTests} corresponds to the images obtained without fake spiral injection.

The PCA-SADI images with $n_{\rm pc}^{\rm ADI} = 1$ and $n_{\rm pc}^{\rm SDI} = 1$ recover visually the full trace of the two injected synthetic spirals (Figure~\ref{SpiralTests}e and i).
By contrast, PCA-ASDI filters out parts of the injected spirals. These are indicated by \emph{cyan arrows} in Figure~\ref{SpiralTests}f and j.
In particular, for both spirals, the cyan arrows appear roughly in the middle of the portion of the spiral where the pitch angle is roughly zero (i.e.~where the arm is most similar to a circle).
We also tested other values of $n_{\rm pc}^{\rm ADI}$ than shown in Figure~\ref{SpiralTests} and noticed that PCA-ASDI filters out spiral \#1 at the location of the cyan arrow in Figure~\ref{SpiralTests}f for all values of $n_{\rm pc}^{\rm ADI}$ from 1 to 10, while spiral \#2 is filtered out at the location of the cyan arrow in~\ref{SpiralTests}j for $n_{\rm pc}^{\rm ADI} \ge 3$.
By contrast, for all values of $n_{\rm pc}^{\rm ADI}$ from 1 to 10, PCA-SADI filters out neither spiral \#1 nor spiral \#2. The only effect of increasing $n_{\rm pc}^{\rm ADI}$ in PCA-SADI is to enhance self-subtraction relatively homogeneously along the spiral trace, so that the full shape is preserved.
%In particular, the negative ADI artefact associated to the bright edge of the outer disc systematically truncates the spiral in aggressive reductions that enable to detect the companion candidate.
Compared to PCA-SADI and PCA-ASDI, PCA-ADI alone does not appear to optimally subtract speckles, as only parts of the synthetic spirals coincident with a bright speckle or the forward-scattered disc edge produce enhanced blobs in the final image (see Figure ~\ref{SpiralTests}g).
%Interestingly, panel (i) of Figure~\ref{SpiralTests} shows a bright blob in the arc at the location where the synthetic spiral meets the arc, similar to the blob observed slightly souther along the arc, where the tentative true spiral meets the arc. The similarity between these two features is another possible indication of the authenticity of the observed spiral-like feature.

The results of this test are in agreement with the survival of extended features identified in our final PCA-SADI image (Figure \ref{PCA-ASDI_final_img}a and c) when increasing $n_{\rm pc}^{\rm ADI}$, while not being recovered in the PCA-ASDI images (Figure~\ref{npcADI_evo}).
We also noticed that PCA-SADI with $n_{\rm pc}^{\rm ADI} = 1$, $n_{\rm pc}^{\rm SDI} = 1$ does not appear to introduce significant geometric biases, as the shape of the retrieved spirals is very similar to that of the injected ones.
This suggests the extended features identified in our final PCA-SADI image are real.

\subsubsection{Surface brightness and throughput}

\begin{figure}
	\centering
	\includegraphics[width=0.48\textwidth]{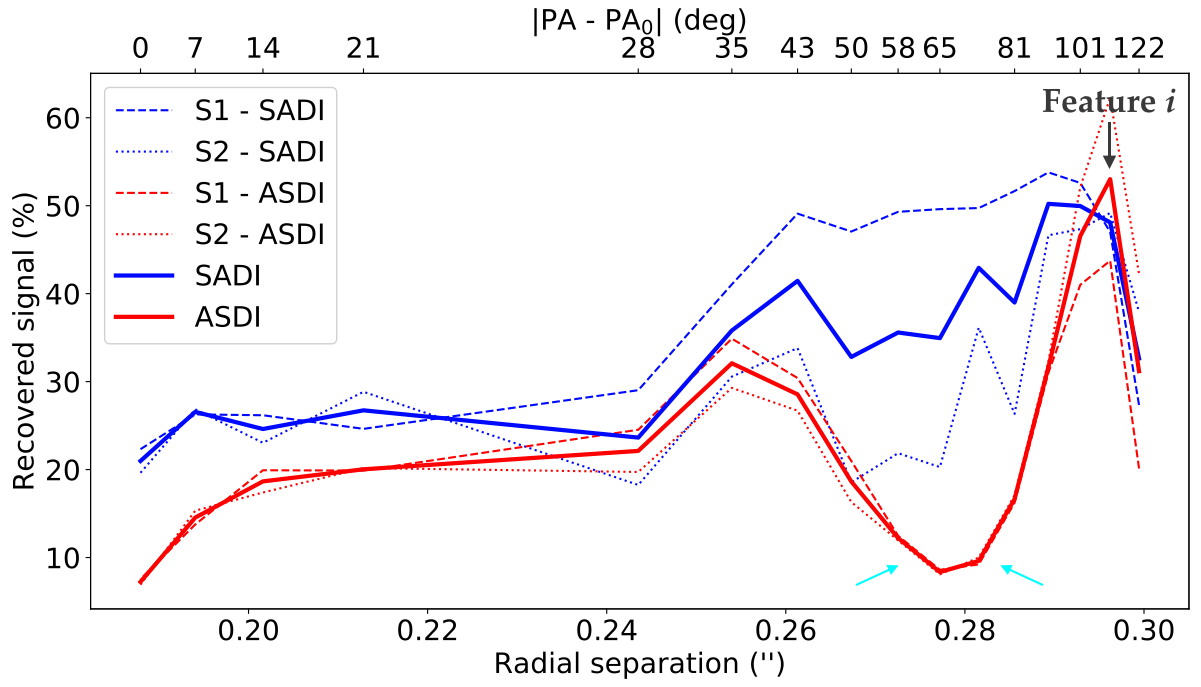}
    \caption{Measured throughput along the two injected spirals with PCA-SADI (\emph{blue curve}) and PCA-ASDI (\emph{red curve}). \emph{Solid curves} correspond to the mean of the individual curves for S1 and S2 (\emph{dashed} and \emph{dotted}, resp.). %As a comparison, the spirals were injected at a constant surface brightness of 77.3 mJy/arcsec$^2$. 
    We note that the spirals are recovered at higher throughput with PCA-SADI, and that the PCA-ASDI profiles present a significant dip likely due to the negative ADI side lobe associated to Feature~\emph{i} (shown with \emph{cyan arrows} as in Figure~\ref{SpiralTests}f and j).
    }
    \label{SB_profile}
\end{figure}

Next, we quantify the amount of signal that is preserved by our algorithms (i.e.~the \emph{throughput}) along each injected spiral. The latter are injected %as a series of consecutive PSFs in each spectral channel and separated in such a way that the spirals have 
at a constant surface brightness of 9.84 mag/arcsec$^2$ (77.3 mJy/arcsec$^2$ at $K$ band) throughout their full trace (Figure~\ref{SpiralTests}d and h). %Using this knowledge, the retrieved surface brightness of the spirals in the final images (Figure~\ref{SpiralTests}e, f, i and j) allows us to estimate the amount of self-subtraction as a function of radius for PCA-SADI and PCA-ASDI. 
Figure~\ref{SB_profile} shows the throughput profile measured in the final PCA-SADI and PCA-ASDI images (\emph{blue} and \emph{red curves}, resp.) along the spirals. Since the injected spirals overlap with authentic extended disc features, %. Therefore, in order to evaluate the surface brightness of the spirals alone, 
we subtracted the flux measured along the spiral traces in the images obtained without spiral injection (i.e.~Figure~\ref{SpiralTests}a and b for PCA-SADI and PCA-ASDI resp.) to that measured in the images obtained after spiral injection (Figure~\ref{SpiralTests}e, f, i and j) before computing the throughput.

%The contrast of the retrieved spirals is shown in Figure~\ref{SB_profile}. %It is given in terms of the contrast of a single of the PSFs used in the series of injection, and compared to the original contrast used for PSF  injection (\emph{dotted line}).
We notice that both algorithms lead to low throughput. Values $\lesssim$50\% are obtained at all radial separations smaller than $\sim 0\farcs3$, with a decreasing trend from large to small radii. This is not surprising: our wavelength range choice (1.95--2.45$\mu$m) leads to insufficient radial motion during the PCA-SDI part of the processing such that partial self-subtraction is unavoidable. Nonetheless, PCA-SADI appears to consistently lead to a higher throughput at all radial separations than PCA-ASDI. Furthermore, the PCA-ASDI profiles show a significant dip at $\sim 0\farcs28$ radius, which is consistent with the dips that were visually identified in Figure~\ref{SpiralTests}f and j (\emph{cyan} arrows).

%Figure*** also allows us to determine the \emph{inner working angle} of the PCA-SADI technique applied to $K$-band SINFONI data: . By definition, it is the radial separation at which the algorithm recovers only 50\% of authentic circumstellar signal. ***

The spiral injection test also allows us to infer the sensitivity achieved by PCA-SADI towards extended structures in our dataset. The injected spirals are indeed rotated/mirror versions of Feature~\emph{iii}, which is identified in Figure~\ref{PCA-ASDI_final_img}c at roughly the limit corresponding to significant signals. We reach thus a sensitivity of $\sim 77$ mJy/arcsec$^2$ in $K$-band total intensity, at $\sim$0\farcs25 separation. % is comparable to the sensitivity of $\sim$0.2 mJy/arcsec$^2$ achieved with HiCIAO in $H$-band polarized intensity beyond $\sim 0\farcs25$ \citep{Hashimoto2012}.
In comparison, \citet{Hashimoto2012} report a sensitivity of $\sim$0.2 mJy/arcsec$^2$ in $H$-band polarized intensity beyond $\sim 0\farcs25$, and a polarization fraction of $\sim 0.5$\%. %Considering the different wavelength, 
%Considering that (1) the polarization fraction is $\sim 0.5$\% of the total intensity at $H$ band \citep{Hashimoto2012} and (2) the scattering cross-section of very small grains decreases quickly with wavelength ($\propto \lambda^{-4}$ for Rayleigh scattering), 
The sensitivity of PCA-SADI towards extended features in our SINFONI data is thus only slightly lower to what is obtained with polarimetric differential imaging \citep[PDI; e.g.][]{Kuhn2001,Quanz2011}, although PDI observations at the same wavelength are required for a better comparison.

Finally, we note that at the tip of spiral~\#1, where the latter intersects the edge of the outer disc, the surface brightness measured in Figure~\ref{SpiralTests}e is twice the value measured in Figure~\ref{SpiralTests}a. This suggests that Feature~\emph{i} has a similar surface brightness of $\sim 77$ mJy/arcsec$^2$ as the injected spirals. %As a comparison, \citet{Hashimoto2012} measured a polarized intensity of $\sim$ 1 mJy/arcsec$^2$ in $H$ band. 
Assuming that Rayleigh scattering is dominant for very small grains, the cross-section is $\propto \lambda^{-4}$ and hence our $K$-band measurement would lead to a prediction of $\sim (1.67/2.18)^{-4} \times 77 \approx 224$ mJy/arcsec$^2$ in $H$-band total intensity. This agrees with the $H$-band polarized intensity of  $\sim$ $1\pm0.2$ mJy/arcsec$^2$ and polarization fraction of $\sim 0.5$\% measured in \citet{Hashimoto2012}.

\begin{figure*}
	\centering
	\includegraphics[width=\textwidth]{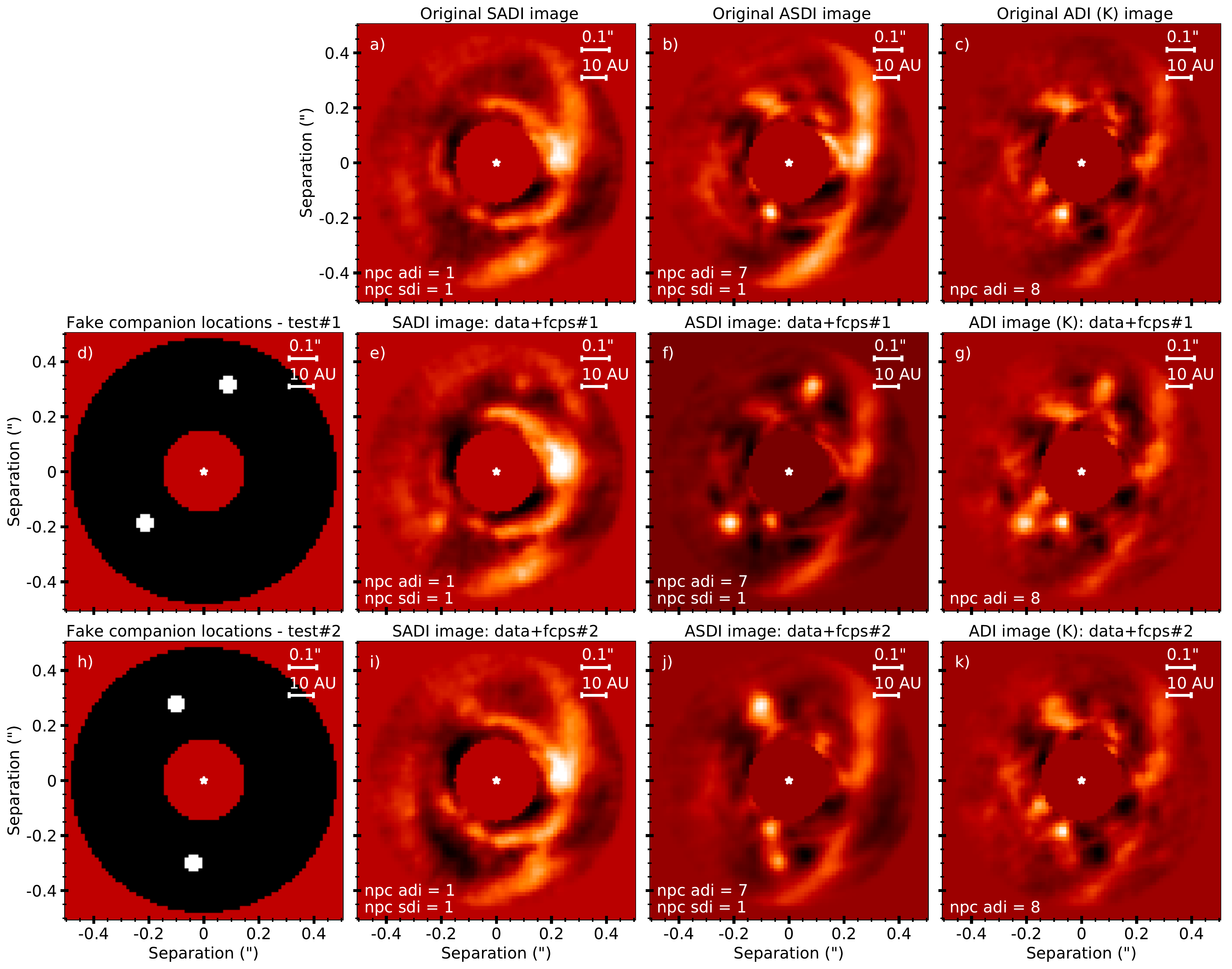}
    \caption{Fake companion injection tests. (\emph{Top row}) Final post-processed images obtained without the injection of any fake companion, for the different algorithms tested in this paper: PCA-SADI, PCA-ASDI and PCA-ADI (K).  (\emph{Middle} and \emph{bottom row}) Images obtained after using the same algorithms, but after injection of fake companions in the original 4D cube, at the locations given in the leftmost column (d and h). In test\#1 (resp.~\#2), the contrast ratio of the injected companions is $\sim 7.7$mag (resp.~$\sim 8.4$mag) in $K$ band, following the spectrum inferred for protoplanet candidate PDS~70~b (Section~\ref{Spectro-photometry}). 
    %({\bf d}) and ({\bf h}): Location of the fake companions injected in the original 4D cube. In test\#1 (resp.~\#2), the contrast ratio of the injected companions is $\sim 7.6$mag (resp.~$\sim 8.4$mag) in $K$ band, following the spectrum inferred for protoplanet candidate PDS~70~b (Section~\ref{Kspectrum}). ({\bf e}), ({\bf f}) and ({\bf g})  PCA-SADI, PCA-ASDI and PCA-ADI (K) images obtained after injection of fake companions of test\#1 in the original 4D cube, respectively. ({\bf i}), ({\bf j}) and ({\bf k})  PCA-SADI, PCA-ASDI and PCA-ADI (K) images obtained after injection of fake companions of test\#1 in the original 4D cube, respectively. %The spiral model used for injection is the best-fit spiral model shown in Figure~\ref{PolarImgs}a, but mirrored along axis PA = 275 \degr. It is fully recovered only in panel d. (\emph{Bottom row}) Final post-processed images obtained after injection of a synthetic spiral injected at the same level as the tentative observed spiral-like feature. Again, 
    In contrast to PCA-SADI, the PCA-ASDI images recover conspicuously all the injected fake companions, suggesting it is more sensitive to point-like sources. %than PCA-SADI. %Flux scales and limits are fixed in each column of images. 
    }
    \label{FcpTests}
\end{figure*}

\subsection{Effect of post-processing on point sources} \label{CompanionInjectionTests}

While the results of the spiral injection tests suggest that PCA-SADI reaches a higher sensitivity for faint extended structures, the fact that only the PCA-ASDI images recover a point-like source at the location of claimed protoplanet candidate PDS~70~b raises the question of whether it corresponds to a filtered extended structure, or whether PCA-ASDI is simply better at recovering point-like sources than PCA-SADI.
In order to answer this question, we injected two pairs of fake companions in copies of our original 4D cube (Figure~\ref{FcpTests}d and h), and post-processed them with PCA-SADI, PCA-ASDI and PCA-ADI (on the collapsed $K$-band channels).
For both tests, companions were (retrospectively) injected based on the contrast as a function of wavelength inferred for the protoplanet candidate (Section~\ref{Spectro-photometry}).
%\citet{Rameau2015} showed that the potential of ASDI and SDI to detect companions strongly depended on their spectrum.  
%It is beyond the scope of this paper to test the recovery of other spectra. 
%However, the conclusions drawn here are expected to be qualitatively similar for all red companions, i.e.~emitting significantly more flux at the long end of the $K$ band than at the short wavelength end.
%While the shape of the injected spectrum is the same as inferred for PDS~70~b for both tests, the contrast ratio was varied in each test. 
In Test\#1, the contrast of the injected companions was rigorously identical to that inferred for PDS~70~b ($\sim 7.6$mag contrast across the $K$ band), while in Test\#2 the fake companions were injected at a twice fainter flux in all channels ($\sim 8.4$mag contrast across the $K$ band).
%The fact that the fake companions are recovered more conspicously than PDS~70~b in Figure~\ref{FcpTests}f reflects the fact the achieved contrast level is slightly deeper further away from the star.  than at the separation of the companion.

PCA-ASDI recovered all injected fake companions for all $n_{\rm pc}^{\rm ADI}$ values in Test \#1, and for all $n_{\rm pc}^{\rm ADI} \ge 3$ in Test \#2 (similar to the $n_{\rm pc}^{\rm ADI}$ values required to recover the protoplanet candidate PDS~70~b).
By contrast, PCA-SADI with $n_{\rm pc}^{\rm ADI} = 1$ and $n_{\rm pc}^{\rm SDI} = 1$ is only able to marginally recover both fake companions of Test \#1 (Figure~\ref{FcpTests}e). 
For increasing values of $n_{\rm pc}^{\rm ADI}$, the detections become more significant, although the achieved SNR values are always lower than obtained with PCA-ASDI.
For test \#2, PCA-SADI only marginally recovers the NE fake companion, but does not recover the other fake companion, for all values of $n_{\rm pc}^{\rm ADI}$.
One of the two brighter fake companions (Test \#1) is also recovered by PCA-ADI (Figure~\ref{FcpTests}g), while the other fake companions appear at a similar level as the speckle noise.

ADI is known to produce characteristic negative side lobes alongside point-source detections. 
A hint of these signatures can be seen in Figure~\ref{FcpTests}g.
PCA-ASDI also produces negative side lobes (Figure~\ref{FcpTests}f and j), confirming the importance of the ADI part in that algorithm.
The low significance of the fake companions in the PCA-SADI images does not enable identification of such features. %clearly, if present.

\subsection{Discussion} \label{SADIvsASDI} %\label{DiscuPostprocessing}

%In this section, 
We have explored different post-processing techniques and reduction parameters in order to make best use of the angular and spectral diversity present in our SINFONI data. 
The application of either SDI or ADI had been previously investigated individually on SINFONI data \citep[][resp.]{Thatte2007,Meshkat2015}. 
Our work is the first attempt to use both. %, which was allowed since SINFONI was offered in combination with a pupil-tracking mode in 2014.

We interpret the poorer results obtained with PCA in a single step compared to our two-step algorithm (PCA-SADI and PCA-ASDI) as speckles being significantly more correlated spectrally than temporally. 
Using two steps takes better advantage of the different speckle correlation levels, while a single larger PCA library dilutes the high spectral cross-correlation which enables an efficient speckle subtraction in the PCA-SDI step. 
PCA-SADI and PCA-ASDI appear thus to be the most appropriate methods for SINFONI datasets. 
This conclusion might not apply for extreme-AO fed IFS such as those of SPHERE/IFS, GPI or SCExAO/CHARIS because 1) the better AO system provides more consistent high Strehl ratio images as a function of time, and 2) the IFS configuration is different (lenslet-based vs image-slicer based for SINFONI) so that the spectral correlation of speckles might be different \citep[e.g.][]{Claudi2008,Wolff2014,Brandt2017}.
%We found that our two-steps PCA algorithm (SDI and ADI separately) performed better than a single step PCA. 
Applying our two-step PCA algorithms %with either PCA-SDI first (PCA-SADI) or PCA-ADI first (PCA-ASDI) both 
led to the recovery of features identified in recent observations of the system using extreme-AO instrument SPHERE \citepalias[Figure~\ref{PCA-ASDI_final_img};][]{Keppler2018, Muller2018}. 

%SADI goes deeper
Using PCA-SDI in the first step is expected to better subtract speckles than PCA-ADI because of the stronger spectral correlation of speckles (acquired simultaneously in all different spectral channels) than the temporal correlation throughout the observing sequence. 
The latter is subject to both slowly varying quasi-static speckles and variations in observing conditions \citep[see also][]{Rameau2015}. 
This interpretation is consistent with the fact that only 1 principal component for both the SDI and ADI parts ($n_{\rm pc}^{\rm SDI} = n_{\rm pc}^{\rm ADI} = 1$) is enough for PCA-SADI to identify all extended structures in the disc, while increasing values of either $n_{\rm pc}^{\rm SDI}$ or $n_{\rm pc}^{\rm ADI}$ progressively self-subtract them while not revealing any new feature. 
%Instead, it progressively self-subtracts the structures identified in the $n_{\rm pc}^{\rm SDI} = n_{\rm pc}^{\rm ADI} = 1$ image. 
%We note that the difference in spectral and temporal speckle correlation levels would be less significant for observations obtained either in very good and stable conditions or with extreme-AO systems.
% SADI => preserves better extended structures
By construction, PCA-SDI preserves azimuthally extended structures, while being potentially more harmful for radially extended structures.
%An advantage of modeling the speckle noise with a minimal number of principal components in PCA-SADI is that self-subtraction of authentic extended structures is overall minimized (as testified by Figure~\ref{SpiralTests}e and i).
However, our final reductions only use the $K$-band spectral channels (instead of the whole $H+K$ range), hence the correspondingly lower amount of radial motion of authentic circumstellar signal (disc or planet) during the SDI part might still induce some (radial) self-subtraction.

% ASDI => point source
Performing PCA-ADI first whitens the noise, hence PCA-SDI in the second step is expected to be less efficient. 
Whitening is not perfect though, in particular for small values of $n_{\rm pc}^{\rm ADI}$, as residual speckles still show significant correlation between different spectral channels even after PCA-ADI (e.g.~Figure \ref{NEGFCconcept}). %e.g.~Figures \ref{NEGFCconcept} and \ref{BrGImages}).
%This is illustrated by our final SADI+Br$\gamma$ image reaching only a slightly deeper contrast than ASDI+Br$\gamma$, as measured from residual throughput-corrected noise level in the image (Section~\ref{BrgammaConstraints}). 
As the temporal correlation is not as high as the spectral correlation, a larger $n_{\rm pc}^{\rm ADI}$ is required for PCA-ASDI than for PCA-SADI to build an optimal model of the speckle pattern.

% ASDI => point sources
Post-processing of disc images with ADI is known to introduce geometric biases, filtering out parts of the disc \citep[e.g.][]{Milli2012,Rameau2017}.
PCA-ADI with a large $n_{\rm pc}^{\rm ADI}$ makes this effect worse because it includes more signal from the disc in the principal components \citep[e.g.][]{Pairet2018a}. %*** TO BE REIMPORTED ***
Similar to the results %of \citet[][]{Milli2012} and \citet{Rameau2017} 
obtained with ADI, %median-ADI and PCA-ADI resp., 
our synthetic spiral injection tests show that PCA-ASDI can also filter out significant %the most radially constant 
parts of the spirals (Figure~\ref{SpiralTests}f and j).
This is consistent with the fact that ADI has a more dominant role in PCA-ASDI than in PCA-SADI. %where, in the latter case, the correlated speckle noise is first subtracted in the SDI step.
%This is similar to the results of previous works highlighting the destructive effect of ADI for azimuthally extended structures \citep[e.g.][]{Milli2012,Rameau2017}, and is to be interpreted as ADI having a more predominant role in PCA-ASDI than in PCA-SADI, where the correlated speckle noise is first subtracted in the SDI step.
Nonetheless, while ADI can filter out azimuthally extended features, it can also enhance --- for the same reason --- the detection of azimuthal asymmetries or point sources. 
This is likely the main reason why only our PCA-ASDI images detect conspicuously PDS~70~b.
Our fake companion injections indeed show that the PCA-ASDI algorithm is more sensitive to faint companions than PCA-SADI (Figure~\ref{FcpTests}) --- at the expense of subtracting azimuthally extended disc structures.

% ccl for ASDI only valid for red companions
\citet{Rameau2015} investigated the sensitivity of an algorithm similar to SADI %but on data acquired with only two (dual-band simultaneous) spectral channels
%The tests presented in \citet{Rameau2015} show that the sensitivity of SADI\footnote{They name it \emph{ASDI}, but is equivalent to our \emph{SADI}.} 
to detect faint companions and % depended strongly on their spectrum.
%They 
found that it 
%the potential of SADI to detect companions 
strongly depended on the spectral features of the companion. 
Therefore in our fake companion injection tests, we considered only artificial companions with the same contrast as a function of wavelength as the protoplanet PDS~70~b (as inferred in Section~\ref{Spectro-photometry}).
%\citet{Rameau2015} showed that the potential of ASDI and SDI to detect companions strongly depended on their spectrum.  
It is beyond the scope of this paper to test the recovery of companions with other spectra. 
However, our conclusions are expected to hold, at least qualitatively, for companions emitting significantly more flux at the long wavelength end of the $K$ band than at the short end.

\begin{figure}
	\centering
	\includegraphics[width=0.49\textwidth]{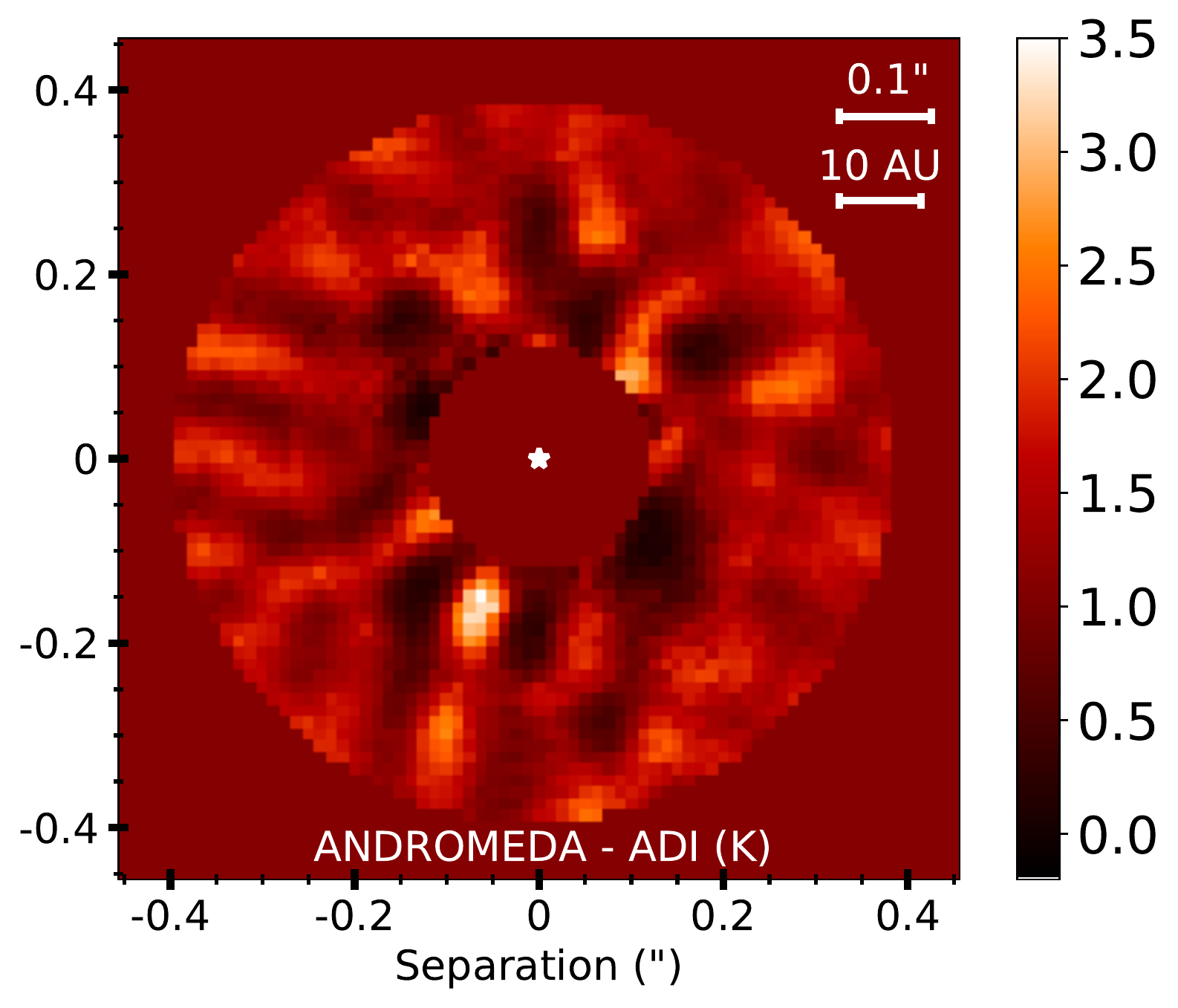}
    \caption{Detection map obtained with ANDROMEDA. PDS~70~b is the only feature found at an SNR $>$ 3.
    }
    \label{DetectionMapANDROMEDA}
\end{figure}

%***In order to better interpret these features and disentangle possible geometric biases inherent to the algorithm, we carried out tests consisting in the post-processing of copies of our original datacube in which we injected either extended features (Section~\ref{SpiralInjectionTests}) or fake companions (Section~\ref{CompanionInjectionTests}).
Our final PCA-ASDI and PCA-SADI images, and the tests carried out in Sections~\ref{SpiralInjectionTests} and \ref{CompanionInjectionTests} alone cannot rule out the possibility that Feature $b$ is the result of filtering of an extended disc feature. 
To confirm the point-source nature of Feature~b, we also applied the ANDROMEDA algorithm \citep[][Cantalloube et al. 2019, in prep.]{Cantalloube2015}.
ANDROMEDA is a maximum likelihood matched-filter algorithm that searches for the expected signature of point-like sources. 
By construction, it is sensitive to authentic point sources, and not to extended disc features filtered by ADI. % into point-like sources.
Figure~\ref{DetectionMapANDROMEDA} shows the detection map obtained after combination of the ANDROMEDA detection maps obtained for all individual spectral channels at wavelength $> 2 \mu$m,
%The final detection map is obtained with 
based on a weighted average proportional to the square of the SNR at each pixel \citep{Thiebaut2016}. 
Feature~b is the only signal detected above $3\sigma$ in our image, providing an independent confirmation of the point-source nature of Feature~b.

\section{Characterization of the protoplanet} \label{CompanionCandidate}

In addition to ANDROMEDA's detection map, several other lines of evidence presented in recent studies argue in favor of Feature~\emph{b} being a protoplanet (see more details in Section~\ref{ProtoplanetRedetection}). %\citep[\citetalias{Keppler2018,Muller2018};][see more details in Section~\ref{ProtoplanetRedetection}]{Wagner2018a}.
Therefore, we attempted to extract both the exact astrometry and the spectrum of Feature $b$.
We followed two different methods: the negative fake companion technique (NEGFC) combined to an adaptation of our PCA-ASDI algorithm (Section~\ref{NEGFC}), and ANDROMEDA (Section~\ref{ANDROMEDAResults}).

\subsection{Spectro-astrometry inferred using NEGFC} \label{NEGFC}

\subsubsection{Negative fake companion technique}
\label{NEGFCresults}

%The results of the tests presented in sections~\ref{SpiralInjectionTests} and \ref{CompanionInjectionTests} suggest that PCA-SADI and PCA-ASDI are more sensitive to azimuthally extended and point-like sources, respectively.
%Therefore, it is possible that both the extended structures and the point-like source identified in Figure~\ref{PCA-ASDI_final_img} co-exist.
%While our final images and tests alone do not rule out the possibility that Feature $b$ is the result of filtering of an extended disc feature

\begin{figure*}
	\centering
	\includegraphics[width=\textwidth]{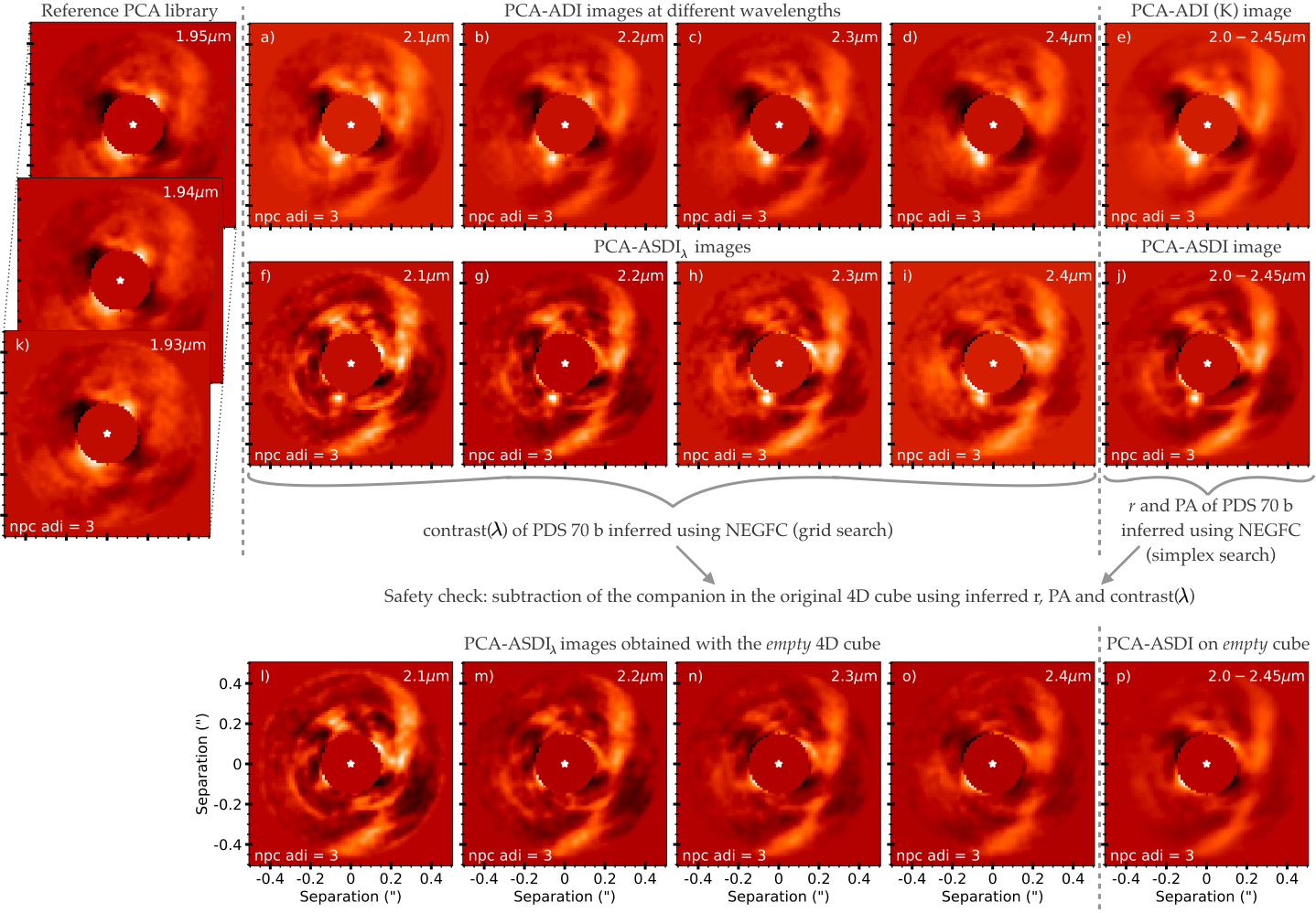}
    \caption{Illustration of the different steps used to infer the contrast as a function of wavelength, radial separation ($r$) and PA of PDS~70~b. ({\bf a}) to ({\bf d}) Median-combined PCA-ADI images for 40 adjacent spectral channels around the considered wavelengths (given in the top right corner of each panel). ({\bf e}) Median-combined PCA-ADI image of all spectral channels of the $K$ band. ({\bf f} to {\bf i}) PCA-ASDI$_{\lambda}$ images and ({\bf j}) PCA-ASDI image, obtained using spectral channels between 1.93 and 1.95 $\mu$m as reference library ({\bf k}) for the PCA-SDI part of the processing. No signal departing from speckle noise can be seen at the location of the companion at those wavelengths. ({\bf l} to {\bf o}) PCA-ASDI$_{\lambda}$ and ({\bf p}) PCA-ASDI images obtained using a copy of our original cube where the companion was subtracted in each spectral channel using contrast($\lambda$), $r$ and PA inferred by the NEGFC algorithm. See Section~\ref{NEGFCresults}.
    %({\bf d}) and ({\bf h}): Location of the fake companions injected in the original 4D cube. In test\#1 (resp.~\#2), the contrast ratio of the injected companions is $\sim 7.6$mag (resp.~$\sim 8.4$mag) in $K$ band, following the spectrum inferred for protoplanet candidate PDS~70~b (Section~\ref{Kspectrum}). ({\bf e}), ({\bf f}) and ({\bf g})  PCA-SADI, PCA-ASDI and PCA-ADI (K) images obtained after injection of fake companions of test\#1 in the original 4D cube, respectively. ({\bf i}), ({\bf j}) and ({\bf k})  PCA-SADI, PCA-ASDI and PCA-ADI (K) images obtained after injection of fake companions of test\#1 in the original 4D cube, respectively. %The spiral model used for injection is the best-fit spiral model shown in Figure~\ref{PolarImgs}a, but mirrored along axis PA = 275 \degr. It is fully recovered only in panel d. (\emph{Bottom row}) Final post-processed images obtained after injection of a synthetic spiral injected at the same level as the tentative observed spiral-like feature. Again, 
    %Contrarily to PCA-SADI, the PCA-ASDI images recover conspicuously all the injected fake companions, suggesting it is more sensitive to point-like sources. %than PCA-SADI. %Flux scales and limits are fixed in each column of images. 
    }
    \label{NEGFCconcept}
\end{figure*}

NEGFC is a specific form of \emph{forward modeling} used to characterize faint point sources, and first proposed in the context of ADI datasets \citep[e.g.][]{Marois2010b,Lagrange2010,Soummer2012}. 
Our adaptation of the NEGFC algorithm is summarized in Figure~\ref{NEGFCconcept}.

%Nonetheless, there are also a number of observational evidence in favor of the protoplanet hypothesis reported in recent studies, including (1) the absence of polarized light at the location of the protoplanet candidate \citep{Keppler2018}, (2) the detection of proper motion consistent with an object on a Keplerian orbit \citep{Muller2018}, (3) a spectrum compatible with models of young substellar objects \citep{Muller2018}, (4) the detection of H$\alpha$ emission \citep{Wagner2018a}, and (5) a tentative velocity dispersion peak in HCO+ at the rough location of the companion candidate \citep{Long2018}.

%\subsubsection{NEGFC results}
%\label{NEGFCresults}

We first attempted to extract the exact position and contrast as a function of wavelength of Feature~\emph{b} using NEGFC coupled with PCA-ADI applied to each spectral channel, as performed for HD~142527~B in \citet{Christiaens2018}.
%Nonetheless, the residual speckle noise in the ADI-processed spectral channels was similar to the flux level of the companion. 
Taking advantage of the whitening of the noise by ADI, we binned (40 by 40) adjacent spectral channels after PCA-ADI processing (Figure~\ref{NEGFCconcept}a to d), and over the whole $K$-band wavelength range (Figure~\ref{NEGFCconcept}e).
%Binning improved the SNR of the companion.
However, we noticed that (1) the pixel values of surrounding bright speckles corresponded to about half of the value of pixels at the location of the companion, and that (2) the contrast estimated by NEGFC (PCA-ADI) corresponded to a roughly twice brighter flux than estimated in the $K1$ and $K2$ filters using SPHERE \citepalias{Muller2018}.
This suggested that a significant contribution from underlying speckles was indeed biasing the flux estimates.

In order to better subtract the residual speckle in the image, PCA-ADI was followed by PCA-SDI, %on the ADI-processed spectral images, 
similar to  the PCA-ASDI algorithm described in Section~\ref{Algorithm}.
PCA-ASDI images (right column of Figure~\ref{npcADI_evo} or Figure~\ref{NEGFCconcept}j) are obtained after PCA-SDI model PSFs are subtracted to each ADI-processed spectral images, and all residual spectral frames of the $K$ band (2.0--2.45$\mu$m) are combined using a variance-weighted average.
In comparison, images labelled PCA-ASDI$_{\lambda}$ (Figure~\ref{NEGFCconcept}f to i) are obtained by combining residual adjacent spectral channels 40 by 40.
We notice in Figure~\ref{NEGFCconcept}f to j that the residual speckle noise has been attenuated compared to images obtained with PCA-ADI alone, so that the signal from the companion appears now to stand out more conspicuously.

Since the spectrum inferred by NEGFC (PCA-ASDI) can be biased if the companion is also present in the PCA library used for the PCA-SDI part \citep[][]{Maire2014,Rameau2015,Galicher2018}, 
we carefully selected the spectral channels to be included in the PCA-SDI library as those with: (1) no signal departing from speckle noise at the location of the companion, and (2) showing the most correlated residual speckle noise with spectral channels where the emission of the companion is detected.
The first condition is met for all spectral channels shortward of $\sim 2.0 \mu$m, while the second condition is only met for $K$-band spectral channels, since the AO correction was significantly poorer in the $H$-band channels.
In order to minimize the risk of self-subtraction, we selected the shortest wavelength spectral channels of the $K$ band that were not significantly affected by telluric line absorption, and hence limited the PCA library to include all spectral channels between 1.93 and 1.95 $\mu$m.
As can be seen in Figure~\ref{NEGFCconcept}k, the residual speckle noise after PCA-ADI in the 1.93--1.95$\mu$m spectral channels correlates with the residual speckle pattern in spectral channels longward of 2.0 $\mu$m (Figure~\ref{NEGFCconcept}a--d), but no significant signal from the companion is detected.
This is consistent with the drop in flux shortward of $\sim$2.1$\mu$m in both the spectrum of the companion and corresponding best-fit models presented in \citetalias{Muller2018}. %, which suggest that the flux of the protoplanet drops below the speckle noise level for channels shortward of $\sim 2.0 \mu$m.

Our strategy is to first use PCA-ASDI to infer the most accurate astrometry of the companion, taking advantage of the enhanced SNR, %stemming from the combination of all $K$-band spectral channels, 
and then to use PCA-ASDI$_{\lambda}$ to extract the contrast as a function of wavelength of the companion, at the exact location inferred with PCA-ASDI.
For this purpose, we adapted the version of NEGFC that is implemented in VIP for PCA-ADI \citep{Wertz2017,GomezGonzalez2017} to work with our PCA-ASDI$_{\lambda}$ and PCA-ASDI implementations. %, and used it to obtain astrometric and spectro-photometric estimates for the companion. 
%This is detailed in the next paragraphs.

\begin{figure*}
	\centering
	\includegraphics[width=\textwidth]{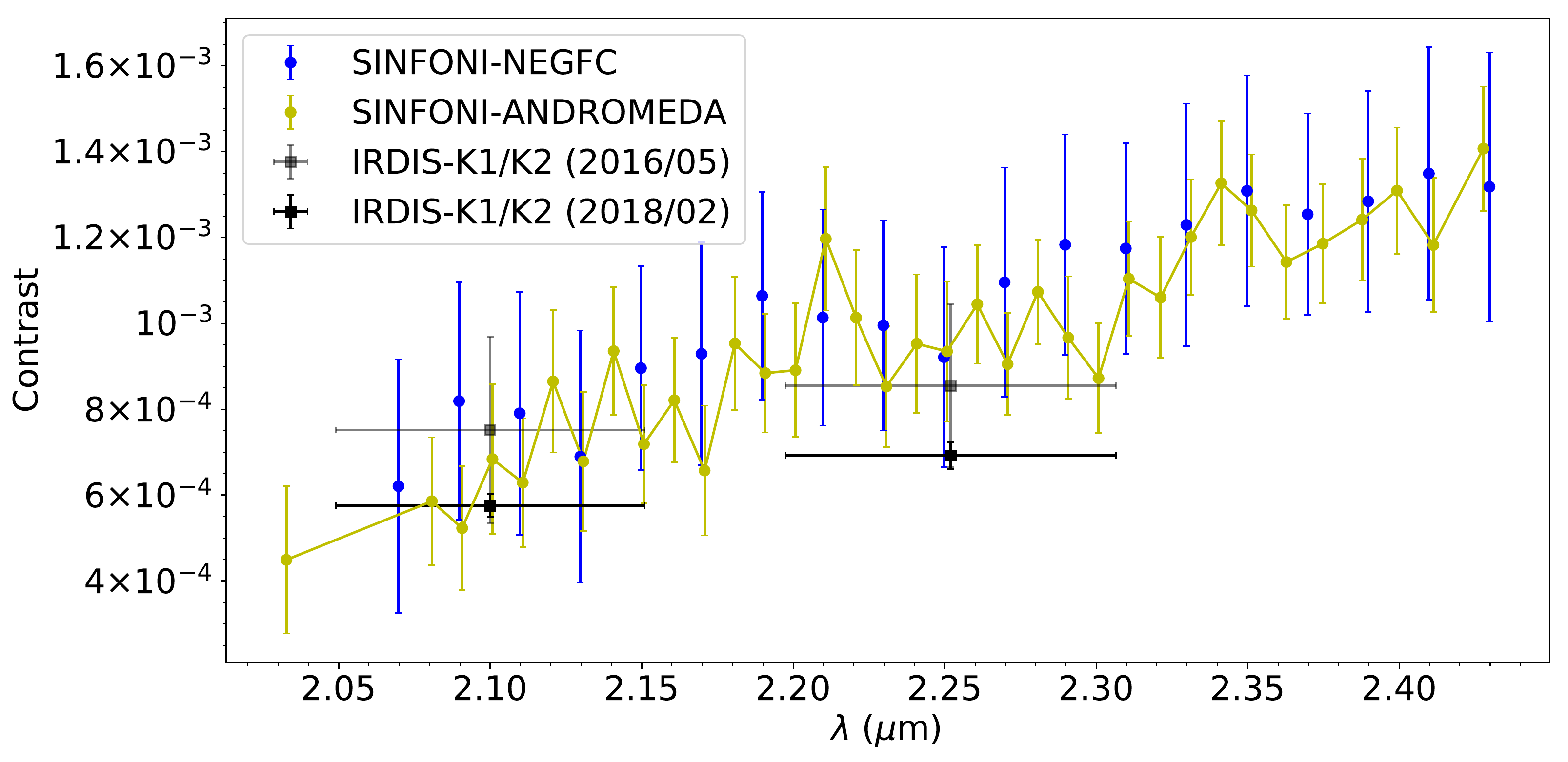}
    \caption{Contrast of the protoplanet with respect to the star as a function of wavelength, as inferred using NEGFC (PCA-ASDI$_{\lambda}$) and ANDROMEDA on our SINFONI data (\emph{blue} and \emph{yellow error bars}, resp.). %, either consisting of photospheric emission alone ({\bf a}) or a combination of photospheric and circumplanetary disc emission ({\bf b}). 
    Both retrieved spectra are consistent with each other, and agree with the broadband measurements obtained with SPHERE/IRDIS in the $K1$ filter on 2016/05 (\emph{grey points}) and 2018/02 (\emph{black points}), and in the $K2$ filter for the first epoch. We note a slight discrepancy with the $K2$ measurement at the second epoch. %(2016/05 in lighter color than the 2018/02 epoch). 
    Horizontal error bars show %represent 
    the FWHM of the $K1$ and $K2$ filters. 
    %\emph{Dashed lines} (BT-SETTL-1 and SB12-1) correspond to the best fit to all points of the spectrum, while \emph{dotted lines} (BT-SETTL-2) only consider points shortward of 1.7 $\mu$m.
    }
    \label{FinalContrast}
\end{figure*}

\subsubsection{Broad-band astrometry}\label{BroadbandAstro-photometry}

In order to obtain the best estimate of the radial separation ($r$) and PA of the companion, we used NEGFC in combination with PCA-ASDI, as it combines all residual spectral frames.
The NEGFC optimization was performed using a Nelder-Mead simplex-based algorithm \citep[see more details in][]{Wertz2017}.
% Figure of merit and aperture used ??!!! 
% Considering larger aperture sizes than 0.6 FWHM leads to significantly higher flux estimates which we interpret as possibly due to contamination from surrounding circumstellar signal.

Instead of considering all spectral channels of the $K$ band for PCA-ASDI, we only consider channels in the [2.13---2.29]$\mu$m wavelength range, as they show less residual speckle noise in the vicinity of the protoplanet. %that appear to otherwise slightly bias the astrometric measurement in other channels.
We noticed indeed that at shorter wavelength than $\sim 2.13 \mu$m, the signal of the companion is too faint, while at longer wavelength than $\sim 2.29 \mu$m, a bright extended feature radially inward of the companion appears to slightly bias the centroid estimate, shifting the estimate to about 1 pixel shorter radius. This bright extended feature is roughly shaped as an inner spiral arm stemming from the location of the companion, and is best seen in Figure~\ref{NEGFCconcept}h and i.

Using NEGFC (PCA-ASDI), we find the radial separation and PA of the companion to be: $193.5 \pm 4.9$ mas and $158.7\degr \pm 3.0$\degr. %method used for speckle noise uncertainty: more injected fake companions?}. %2.9mas + 0.2degr of residual speckle noise
Given the PA of the outer disc (PA$\sim$159\degr), both the projected and deprojected physical separations are 20.9$\pm$0.6 au.
The quoted astrometric uncertainties reflect (1) the error associated with stellar centering; (2) instrumental systematic uncertainties (plate scale, true North and pupil offset); and (3) residual speckle noise.
Comparison between centering with 2D Gaussian and 2D Moffat functions are consistent within $\sim$0.05 pixel, and we hence expect the error associated to stellar centering to not be significantly larger than that.
We conservatively consider the same systematic uncertainties as estimated in \citet{Meshkat2015} for SINFONI data obtained at a similar epoch, but for a companion at larger separation from its star than PDS~70~b: 0.4~mas and 0.5\degr. 
The error associated to residual speckle noise dominate the budget: 4.8~mas and 2.9\degr, for $r$ and PA respectively. %***CHANGE FOR 42 ***
The procedure used to estimate them is similar to that presented in \citet{Wertz2017} and is detailed in appendix~\ref{tests_spectrophotometry}.

\subsubsection{Spectro-photometry}\label{Spectro-photometry}

We applied NEGFC (PCA-ASDI$_{\lambda}$) to determine the contrast of the companion as a function of wavelength.
The position of the companion was fixed to that inferred in Section~\ref{BroadbandAstro-photometry} using NEGFC (PCA-ASDI), leaving only a single free parameter to explore.
A grid search was performed to infer the optimal contrast of the companion in each binned spectral channel.
Given that the star was not saturated in our datacube, the centering of the frames based on fitting the stellar centroid with a 2D Moffat function is expected to be very accurate.
Therefore, the position of the companion is not expected to vary significantly in our post-processed spectral channels.

%The main reasons for 
We did not use the Nelder-Mead simplex-based algorithm on the 3 parameters together in each (binned) spectral channel for the same reasons we limited NEGFC (PCA-ASDI) to spectral channels between 2.13--2.29 $\mu$m for the astrometric estimate in Section~\ref{BroadbandAstro-photometry}. %is the relatively low SNR of the companion in spectral channels between $\sim$ 2.0 and 2.15 $\mu$m.
We noticed that for some of the spectral channels outside of that wavelength range, the Nelder-Mead algorithm converged at erroneous locations, %likely due to neighbouring residual speckles of comparatively significant level, 
hence biasing the inferred spectro-photometry of the companion.
The grid search is also faster, which facilitated the estimation of both the residual speckle uncertainties and the optimal $n_{\rm pc}^{\rm ADI}$ used for contrast estimation in each binned spectral channel (appendix~\ref{tests_spectrophotometry}).
%We noticed that the detection was most conspicuous between 2.22 and 2.38 $\mu$m, which is consistent will smaller uncertainties inferred in that wavelength range.
The final inferred contrast($\lambda$) is shown with blue error bars in Figure~\ref{FinalContrast}.

\subsection{Spectro-photometry inferred using ANDROMEDA} \label{ANDROMEDAResults}

%***ASTROMETRY AS WELL? => IF SO, CHANGE TITLE TO "SPECTRO-ASTROMETRY"

Based on individual ANDROMEDA detection maps, we selected the spectral channels where the significance of the protoplanetary signal was larger than 3$\sigma$, and built a spectrum based on the flux estimated by ANDROMEDA. 
These spectral channels were all located in the $K$ band. 
As we noticed the presence of some potential outliers, we median-combined the fluxes inferred by ANDROMEDA in adjacent channels 20 by 20.
The resulting spectrum is shown with \emph{yellow error bars} in Figure~\ref{FinalContrast}.
It is consistent with both the spectrum inferred using NEGFC (Section~\ref{Spectro-photometry}), and the SPHERE/IRDIS measurements acquired in the $K1$ filter ($\sim 2.10 \mu$m) on 2016/05/14 and 2018/02/24, but only the $K2$ measurement ($\sim 2.25 \mu$m) acquired at the first epoch \citepalias{Keppler2018,Muller2018}.
%Both the smaller error bars and ability to retrieve a higher resolution spectrum with ANDROMEDA than with NEGFC convey the better isolation of the point-source signal from extended features with ANDROMEDA compared to PCA-ASDI$_{\lambda}$. 
Most fluxes estimated by either NEGFC or ANDROMEDA between 2.19 and 2.31 $\mu$m (all but two points) are slightly larger than the 2018/02 SPHERE measurement.
A detailed spectral analysis of the protoplanet is beyond the scope of this paper %. %, which is devoted to the analysis of our VLT/SINFONI data alone.
%Spectral characterization of the protoplanet including photometric measurements obtained at other wavelengths 
and is deferred to a forthcoming paper (Christiaens et al.~2019, submitted to ApJL).

We note that ANDROMEDA performs better than NEGFC+PCA-ASDI for the extraction of the spectrum of the companion. It leads to smaller uncertainties and preserves a finer spectral resolution. This is because ANDROMEDA is not sensitive to extended disc features, while PCA-ASDI does not entirely filter out extended signals from the disc, which complicates the estimation of the flux stemming from the point source only.

\begin{figure*}
	\centering
	\includegraphics[width=\textwidth]{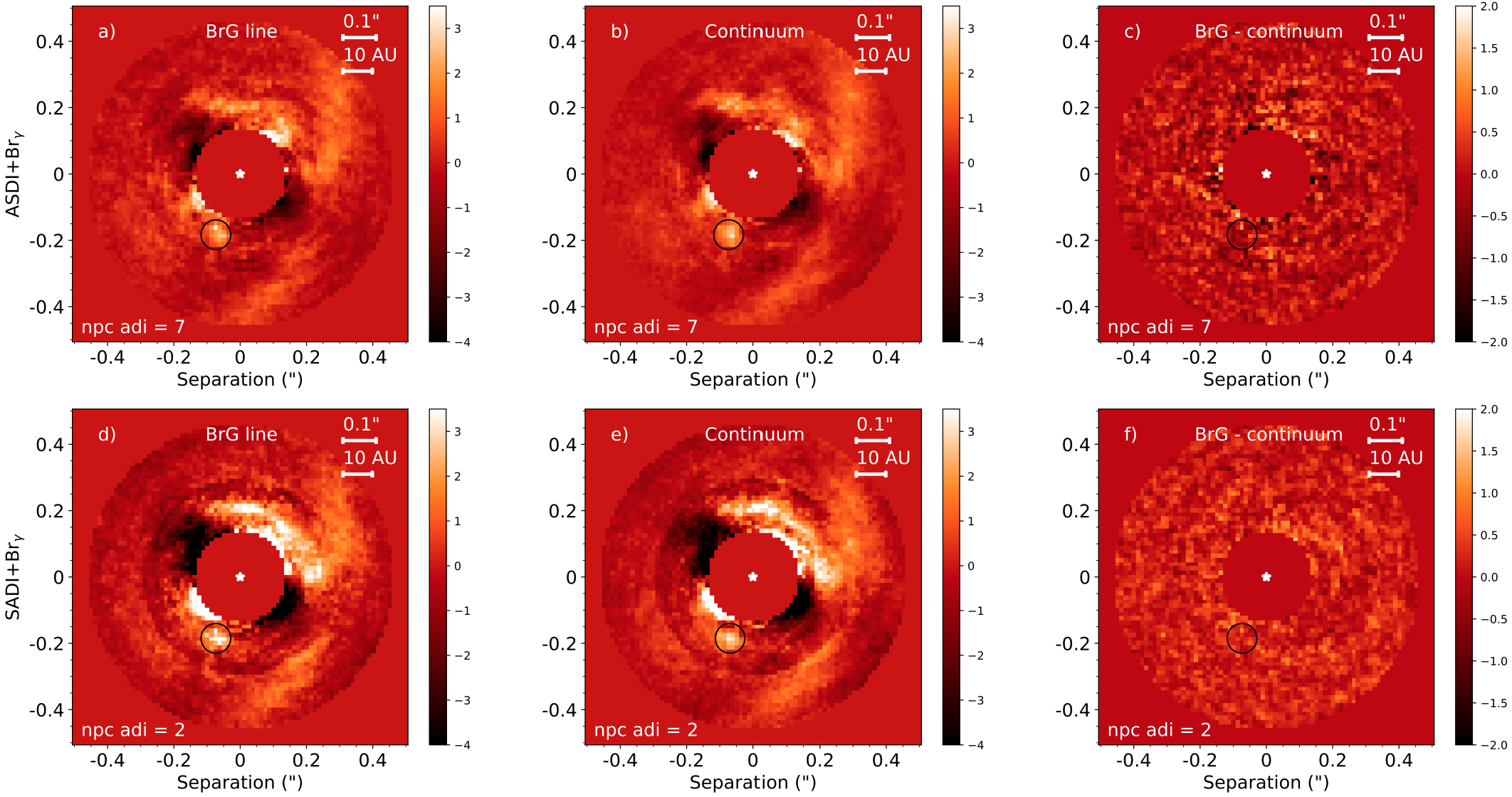}
    \caption{Results of \emph{ASDI+Br}$\gamma$ (first row) and \emph{SADI+Br}$\gamma$ (second row). {\bf a)} and {\bf d)}: weighted average of the PCA-ADI images obtained with the Br$\gamma$ line channels, %($I_{\textrm{Br}\gamma}$), 
 {\bf b)} and {\bf e)}: median of the PCA-ADI images obtained with the continuum line channels, %$I_{\textrm{cont}}$, 
 {\bf c)} and {\bf f)} final \emph{ASDI+Br}$\gamma$ and \emph{SADI+Br}$\gamma$ images, obtained with the respective optimal number of principal components (resp. $n_{\rm pc}^{\rm ADI}  = 7$ and $n_{\rm pc}^{\rm ADI}  = 2$). See Appendix~\ref{AppBrG} for details on each algorithm. %, obtained as $I_{\textrm{Br}\gamma} - I_{\textrm{cont}}$. 
 No significant signal is left at the location of the companion (indicated by a black circle in all panels), nor elsewhere in the field, in the \emph{ASDI+Br}$\gamma$ and \emph{SADI+Br}$\gamma$ images.
 }
    \label{BrGImages}
\end{figure*}

\subsection{Constraints on Br$\gamma$ emission} \label{BrgammaConstraints}

\citet{Wagner2018a} recently reported the tentative detection of H$\alpha$ emission originating from the location of the protoplanet candidate.
In this section, we investigate the presence of Br$\gamma$ line emission (2.16612 $\mu$m) at the location of the companion.
We applied two slightly different algorithms for detection of Br$\gamma$ emission, referred to as \emph{SADI+Br}$\gamma$ and \emph{ASDI+Br}$\gamma$, and that we adapted from our PCA-SADI and PCA-ASDI algorithms respectively. % to avoid confusion with PCA-SADI and PCA-ASDI presented in the previous sections. 
The two techniques are described in details in Appendix~\ref{AppBrG}. %the following paragraphs.
%\emph{ASDI+Br}$\gamma$ is similar to what is referred to as \emph{ASDI} in \citet{Close2014} and \emph{SDI+} in \citet{Wagner2018a}.

Neither the final \emph{ASDI+Br}$\gamma$ image nor the final \emph{SADI+Br}$\gamma$ image (Figure~\ref{BrGImages}c and f, resp.) reveal any significant Br$\gamma$ signal, neither at the location of the companion nor elsewhere in the image. 
We note that the throughput-corrected residual noise is slightly lower in the final $I_{\textrm{SADI+Br}\gamma}$ image than in the $I_{\textrm{ASDI+Br}\gamma}$ image.
%This is in line with PCA-SADI reaching slightly better contrast than PCA-ASDI
%CONTINUE HERE
We hence focus on the former one in the rest of this section to estimate an upper limit on the Br$\gamma$-line flux of protoplanet candidate PDS 70 b.
%The throughput-corrected residual noise is slightly lower in the final $I_{\textrm{SADI+Br}\gamma}$ image than in the $I_{\textrm{ASDI+Br}\gamma}$ image, and we hence focus on the former one.
The hypothetical flux that would be measured in $I_{\textrm{SADI+Br}\gamma}$ at the location of the protoplanet candidate can be expressed as follows:
\begin{align}
\label{Eq:MeasFlux}
%F_{\rm meas} = \big[ F_{\textrm{Br}\gamma, p} + F_{\textrm{cont}, p} + \epsilon_{\textrm{Br}\gamma, \star}\big]_{I_{\textrm{Br}\gamma}} - \big[ \frac{F_{\textrm{Br}\gamma, \star}}{F_{\textrm{cont}, \star}} F_{\textrm{cont}, p} + \epsilon_{\textrm{cont}, \star}\big]_{I_{\textrm{cont}}} % TOO LONG!
F_{\rm meas} = \big[ F_{\textrm{Br}\gamma, p} + F_{\textrm{cont}, p} + \epsilon_{\textrm{Br}\gamma, \star}\big] - \big[ \frac{F_{\textrm{Br}\gamma, \star}}{F_{\textrm{cont}, \star}} F_{\textrm{cont}, p} + \epsilon_{\textrm{cont}, \star}\big]
\end{align}
where $F_{\textrm{Br}\gamma, p}$ and $F_{\textrm{cont}, p}$ are the Br$\gamma$ line and continuum fluxes of the planet, $F_{\textrm{Br}\gamma, \star}$ and $F_{\textrm{cont}, \star}$ are the Br$\gamma$ line and continuum fluxes of the star, and $\epsilon_{\textrm{Br}\gamma, \star}$ and $\epsilon_{\textrm{cont}, \star}$ are the contribution of residual speckle signal at the location of the planet in Br$\gamma$ line and continuum channels, respectively.
Fraction $F_{\textrm{Br}\gamma, \star}$/$F_{\textrm{cont}, \star}$ stems from the rescaling of continuum channels performed at the beginning of the \emph{SADI+Br}$\gamma$ procedure.

The residual speckle noise in $I_{\textrm{Br}\gamma}$ and $I_{\textrm{cont}}$ appears well correlated (Figure~\ref{BrGImages}d and e), which is also confirmed by the absence of residual signal at the spatial frequency expected for speckles (size $\approx$ FWHM) in $I_{\textrm{SADI+Br}\gamma}$ (Figure~\ref{BrGImages}f). 
Therefore we can consider $\epsilon_{\textrm{Br}\gamma, \star}$ and $\epsilon_{\textrm{cont}, \star}$ in Eq.~\ref{Eq:MeasFlux}) to efficiently cancel out.
Rewriting Eq.~\ref{Eq:MeasFlux} in terms of Br$\gamma$ line flux of the protoplanet candidate yields:
\begin{align}
\label{Eq:FluxBrG}
F_{\textrm{Br}\gamma, p} = F_{\rm meas} - \bigg(1-\frac{F_{\textrm{Br}\gamma, \star}}{F_{\textrm{cont}, \star}}\bigg) F_{\textrm{cont}, p}% + \epsilon_{\star}
\end{align}
Given the absence of significant signal at the location of the companion in Figure~\ref{BrGImages}f, we replace $F_{\rm meas}$ by an upper limit corresponding to 5 times the noise level at the radial separation of the companion. %: $1.1 \times 10^{-19}$ W m$^{-2}$ (considering the 7.5 \AA~FWHM). 
This noise level is estimated from the standard deviation of the fluxes measured in independent apertures at that radial separation, corrected by a factor accounting for small sample statistics \citep{Mawet2014}.
Given the efficient speckle removal and whitening of the noise, we consider that 5 times this noise level is a conservative estimate of the 5$\sigma$ contrast reached at the separation of the companion.
It corresponds to $\sim 6.3 \times 10^{-4}$ ($\sim$ 8.0 mag) contrast.
An estimation for $F_{\textrm{cont}, p}$ can be obtained from the photometric point directly longward to the Br$\gamma$ line in Figure~\ref{FinalContrast}: $\sim$7.6~mag contrast.
%Considering a 7.5 \AA~FWHM, $F_{\textrm{cont}, p} \approx 1.6 \times 10^{-19} \pm ***$ W m$^{-2}$.  ***FINAL VALUE TO BE UPDATED***
Since $F_{\textrm{Br}\gamma, \star}/F_{\textrm{cont}, \star} \approx 0.99$, the second term of Eq.~\ref{Eq:FluxBrG} is thus almost two orders of magnitude smaller than our upper limit on $F_{\rm meas}$, and can be neglected.

% Second term is -1e-18 W m-2 um-1, which is smaller than our 5sigma uncertainty on F_meas: 1e-16 W m-2 um-1
% 
%Residual shot noise is estimated from the standard deviation of the fluxes measured in independent 1FWHM-wide apertures at the radial separation of the companion.
%We consider that 5 times the level of that residual shot noise, corrected for small sample statistics \citep{Mawet2014}, is a conservative estimate of the 5$\sigma$ contrast reached at the separation of the companion.
%It corresponds to $\sim 7 \times 10^{-4}$ ($\sim$ 8.0 mag) *** which is close to the estimated flux of the companion in the adjacent continuum ***Previous Sec.
%The error on the absolute value of $F_{\textrm{Br}\gamma, p}$ is also directly related to our uncertainty on $F_{\textrm{cont}, p}$ scaled by $1-F_{\textrm{Br}\gamma, \star}/F_{\textrm{cont}, \star}$.
%We measure in our original spectral cube (before ASDI processing) that $F_{\textrm{Br}\gamma, \star}/F_{\textrm{cont}, \star} \approx 0.99$.
%Therefore, the uncertainty associated to that term is about 2 orders of magnitude lower than that associated to residual shot noise, and is neglected in comparison.
%Our final uncertainty on the value of $F_{\textrm{Br}\gamma, p}$ is obtained by summing in quadrature the uncertainties related to the residual shot noise and the continuum flux of the companion.
Our final upper limit on $F_{\textrm{Br}\gamma, p}$ is $\sim 8.3 \times 10^{-20}$ W m$^{-2}$. %OLD WITH ASDI: 1.1 \times 10^{-19}$ W m$^{-2}$. 
It is obtained conservatively considering the 5$\sigma$ contrast achieved in our final $I_{\textrm{SADI+Br}\gamma}$ image, the calibrated spectrum of the star reported in \citet{Long2018} and a 7.5\AA -FWHM linewidth. %-1nm equivalent width \citep[e.g.][]{Mendigutia2014}.
Considering a distance of 113~pc, this translates into a Br$\gamma$ luminosity $\log(L(\mathrm{Br}\gamma / L_{\odot}) < -7.48$. %OLD: -7.35
\citet{Calvet2004} inferred the following empirical relationship to convert $L(\mathrm{Br}\gamma)$ into an accretion luminosity:
\begin{equation}
\log (L_{\mathrm{acc}}/ L_{\sun}) = 0.9 \times (\log L(\mathrm{Br}_{\gamma}) / L_{\sun} + 4) -0.7.
\end{equation}
Although this expression was inferred for T-Tauri stars, we apply it to PDS~70~b due to the lack of similar relationship for lower mass objects.
 We find $ \log (L_{\mathrm{acc}}/ L_{\sun}) < -3.83$. % OLD: -3.71
We convert this accretion luminosity to a mass accretion rate using $\dot{M_b} = 1.25 L_{\mathrm{acc}} R_b/G M_b$ \citep{Gullbring1998}, and find: $\dot{M_b} < 1.26 \times 10^{-7} \big[ \frac{5 M_{\rm Jup}}{M_b} \big] \big[ \frac{R_b}{R_{\rm Jup}}\big] M_{\rm Jup} $ yr$^{-1}$. 

\subsection{Discussion} \label{ProtoplanetRedetection}

%\subsection{Protoplanet redetection}\label{ProtoplanetRedetection}

Previous claims of protoplanet PDS~70~b detections all made use of ADI \citep[\citetalias{Keppler2018,Muller2018};][]{Wagner2018a}.
Considering our PCA-SADI image and our fake spiral injection tests alone, it is not possible to discard the possibility that Feature~\emph{b} is tracing an extended disc feature that is filtered by ADI. 
Given the past investigations on the effect of ADI on discs \citep[e.g.][]{Esposito2014,Follette2017}, including the blobs created along the semi-major axis of thin inclined rings \citep{Milli2012}), our results would actually suggest extreme caution regarding the blob located along the semi-major axis of the transition disc of PDS~70.
The location of Feature~\emph{b} at the tip of a spiral-like feature (Feature~\emph{iii}) seen in the PCA-SADI image is also reminiscent of HD~100546, for which \citet{Rameau2017} showed that a spiral seen in RDI images could produce a point source similar to protoplanet candidate HD~100546~b after ADI filtering \citep{Quanz2013}.
%Since the point-source, {\bf referred to as PDS~70~\emph{b} in \citet{Keppler2018, Muller2018, Wagner2018a}}, is not recovered in that reduction, but rather shows up at the tip of an extended spiral-like feature after aggressive processing, we cannot rule out the possibility that it is in fact a filtered portion of the spiral, as was suggested for HD~100546~b \citep{Follette2017,Rameau2017}.

However, in contrast to other protoplanet candidates, several lines of evidence argue in favor of the protoplanet hypothesis in the case of PDS~70~b, apart from the point source detection:
\begin{enumerate}
\item the proper motion of the protoplanet is consistent with an object on a Keplerian orbit \citepalias{Muller2018}; 
\item the inferred spectrum of the protoplanet is compatible with synthetic spectral models of young substellar objects \citepalias{Muller2018}; 
%\item no polarized light is detected at the location of the companion \citepalias{Keppler2018};
\item H$\alpha$ emission is tentatively detected at the protoplanet location \citep{Wagner2018a}; and 
%\item a tentative velocity dispersion peak is seen in the ALMA HCO$^+$ J=4-3 velocity dispersion map at the rough location of the companion candidate \citep{Long2018}, which is the kinematic imprint roughly expected from a $\sim 5 M_{\rm Jup}$ planet \citep{Perez2015}.
\end{enumerate}
These arguments motivated us to extract the exact astrometry and spectro-photometry of Feature~\emph{b} (Section~\ref{NEGFCresults}).
We found the position of the companion ($r = 193.5 \pm 4.9$mas and PA $= 158.7\degr \pm 3.0$\degr) to be consistent with detections at other epochs \citepalias{Keppler2018,Muller2018}.
The contrast as a function of wavelength is qualitatively consistent with the two photometric points inferred with SPHERE/IRDIS in the $K1$ and $K2$ filters.
Our analysis thus further strengthens points (i) and (ii) listed above.
A detailed spectral analysis of the companion using both our SINFONI spectrum and the rest of the SED presented in \citetalias{Muller2018} is beyond the scope of this paper, and will be presented in a forthcoming paper.

An alternative method to spectrally characterize the companion is to cross-correlate spectral templates \citep[e.g.][]{Konopacky2013,Hoeijmakers2018}.
The application of this promising technique is beyond the scope of this work, but could provide further spectral information on the protoplanet. 
%AO-fed IFUs like VLT/SINFONI or Keck/OSIRIS are great to perform high contrast imaging and characterization at medium spectral resolution; not available behind extreme-AO systems yet. 

%***DEPENDING ON FINAL SINFONI ERROR BARS: discuss possibility of variation in flux of the companion at K2 wavelength - brighter at SINFONI epoch?***

%Therefore, I attempted to extract the K-band spectrum of the point-like source, using a tweaked version of NEGFC+PCA-ASDI (concept explained in Fig. 5). The quality of the extracted spectrum reached beyond my expectations. The spectrum is qualitatively consistent with the 2 photometric points inferred with SPHERE/IRDIS, but its shape is best fit by the combination of both atmospheric (BT-SETTL) and circumplanetary disc emission (cf. models presented in Zhu+15). 

%\subsubsection{Accretion rate of PDS~70~b and extinction}

%In Section~\ref{BrgammaConstraints}, we slightly modified our PCA-ASDI and PCA-SADI algorithms to search for Br$\gamma$ emission from PDS~70~b and, based on our non-detection, 
Based on the non-detection of Br$\gamma$ emission at the location of the companion, we constrained its accretion rate to be $\dot{M_b} < 1.26 \times 10^{-7} \big[ \frac{5 M_{\rm Jup}}{M_b} \big] \big[ \frac{R_b}{R_{\rm Jup}}\big] M_{\rm Jup} $ yr$^{-1}$. 
A major uncertainty in this value is whether the empirical relationship to convert Br$\gamma$ luminosity to accretion luminosity inferred for T-Tauri stars is still valid in the planetary mass regime \citep{Calvet2004}.
Regardless, our upper limit is compatible with the mass accretion rate inferred from the tentative detection of H$\alpha$ emission and using a similar T-Tauri-based conversion relationship: $10^{-8\pm1} M_{\rm Jup}$ yr$^{-1}$ for a 5--9 $M_{\rm Jup}$ planet \citep{Wagner2018a}.

The uncertainty on the accretion rate measured in \citet{Wagner2018a} stems partially from the unknown value of the extinction in the line of sight of the companion.
For $A_V \sim 3$mag, the measured H$\alpha$ flux of the protoplanet candidate would imply $\dot{M_b} \sim 10^{-7.0} M_{\rm Jup}$ yr$^{-1}$ (for a 5--9 $M_{\rm Jup}$).
Therefore, our upper limit on the accretion rate based on the Br$\gamma$ line -- significantly less affected by extinction -- suggests that the line of sight towards the protoplanet candidate has a visual extinction $A_V < 3$mag.

\section{Characterization of disc features}\label{DiscFeatures}

%\subsection{Recovery of extended signals}

%{\bf REQUESTED TESTS BY REFEREED - TO BE DONE}

%\subsection{Discussion} 
\label{Disc-ussion}
% STIM MAPs
%The STIM map (bottom row of Figure~\ref{PCA-ASDI_final_img}; appendix~\ref{SNR_vs_STIM}) is the most appropriate tool to identify significant signals in our images \citep[][]{Pairet2018}. It bypasses the limitations that are inherent to the classical SNR map: (i) small number statistics at short separations and (ii) biased noise estimates in the case of extended authentic circumstellar signals in the field \citep{Mawet2014}.
Based on the STIM maps of our PCA-SADI and PCA-ASDI images (bottom row of Figure~\ref{PCA-ASDI_final_img}), we identified similar features as in the image obtained from a long integration of PDS~70 using extreme-AO instrument VLT/SPHERE \citepalias{Muller2018}: the bright forward-scattered edge of the disc (Feature~\emph{i}), several azimuthally extended features (Features~\emph{ii},~\emph{iv} and v), the faint back-scattered light from the far side of the disc (Feature vi), and a bright blob (Feature $b$), interpreted as a protoplanet. 
The root of features ii and iv is consistent with the location of the \emph{spur} recently observed in dust continuum and $^{12}$CO emission \citep{Keppler2019}. 
The spur could be tracing the tip of a gap-crossing stream. Such stream would physically require the presence of an additional companion, which could account for spiral-shaped features \emph{ii}/\emph{iv} seen in our images.
%Considering that extended features in our NIR images trace the surface of the flared disc, features ii/iv might correspond to the spiral arm launched by the companion candidate, if the latter is in the midplane of the disc.}

%Features i and vi are tracing the bright forward-scattered edge and faint back-scattered light from the close and far side of the disc, resp., as also confirmed by previous polarized light images of the disc \citep[][\citetalias{Keppler2018}]{Hashimoto2012}.
%Among possible extended features in the cavity, we only focus our discussion on features that are considered significant based on the STIM map of the PCA-SADI image (Figure~\ref{PCA-ASDI_final_img}c; i.e.~features~\emph{ii} and\emph{iii}).
%Feature~\emph{ii} appears to be linked to an overdensity in the forward-scattered edge of the disc, and is similar to a blend of features (2) and (3) identified in \citetalias{Muller2018}.
%It might be tracing a possible gap-crossing stream \citep{Price2018}.

Feature \emph{iii} is the only feature with no clear counterpart in the \citetalias{Muller2018} image. 
It is best seen in our final PCA-SADI image, although a hint of its presence can also be seen in the PCA-ASDI images.
It appears to stem from the location of the protoplanet and could either trace an outer spiral or a gap-crossing stream connected to the outer disc. %, possibly extending into Features~\emph{ii}/iv after distortion by the bright edge of the outer disc.
We estimate the deprojected pitch angle of Feature~iii to be $27\pm4$\degr~based on the best-fit logarithmic spiral, if it lies in the
disc plane.
If authentic, %what would be the possible reasons of its non-detection 
why was it not seen in the long-integration SPHERE/IRDIS observation \citepalias{Muller2018}?
The \citetalias{Muller2018} image was obtained using ADI, which is known to create negative azimuthal side lobes around bright features.
In our synthetic spiral injection tests (Figure~\ref{SpiralTests}), we noticed that a spiral with the same shape as Feature~\emph{iii}  was cancelled by PCA-ASDI in the vicinity of the bright disc edge, where the spiral's pitch angle was also the lowest (close to an arc circle).
This could be due to either self-subtraction and/or hiding by the negative lobe of the bright disc edge.
As pointed out in Section~\ref{SADIvsASDI}, ADI has a dominant role in our PCA-ASDI algorithm. 
It is thus possible that the feature was cancelled out in the \citetalias{Muller2018} image for the same reason.
%We note that the injected spirals are not visible in our PCA-ADI (collapsed $K$-band) images either (Figure~\ref{SpiralTests}), however it is not possible to discard the possibility that this is due to high residual speckle noise level in that case.

%The latter is close to the bright forward-scattered edge of the disc and could be significantly attenuated by the associated ADI negative side lobe, in a similar fashion as is injected spiral \#1 in Figure~\ref{SpiralTests}f.
%***Could be filtered out by ADI in a similar way PCA-ASDI appears to filter out both Features~\emph{iii} and~\emph{iv} identified with PCA-SADI.
%Feature~\emph{iii} appears to stem from the location of the protoplanet.
%It could either trace a gap-crossing stream connected to the outer disc, or an outer spiral that could be related to Features~\emph{ii}/\emph{iv} but appears distorted t

% ARE EXTENDED FEATURES REAL?
Apart from the forward-scattered edge of the outer disc, neither the extended features identified in out SINFONI mages, nor those identified in the median-ADI image of \citetalias{Muller2018}, have a clear counterpart in polarized light images \citepalias{Keppler2018}.
Polarimetric differential imaging \citep[PDI; e.g.][]{Kuhn2001,Quanz2011} is often considered the best method to image extended disc features.
%*** Tentative evidence in PDI images after subtraction of central unresolved polarized signal of residual \citep[Figure~1 in][]{Keppler2018}. 
The absence of extended features identified with ADI or SADI in PDI images of PDS~70 suggests caution.
%fact that PDI images of PDS~70 does not reveal any of the intra-gap extended features identified with ADI or SADI suggests caution.
%Only faint and tentative signals can 
Alternatively, non-optimal observing conditions (mediocre seeing), a shorter integration time and the fact that polarized intensity contains only a fraction of the total intensity, might account for the lack of a conspicuous counterpart in polarized light to the features identified in our images \citepalias{Keppler2018}.
%We note that the intra-gap extended structures identified in the median-ADI image of \citetalias{Muller2018} have no clear counterpart in polarized light images either \citepalias{Keppler2018}.

Based on previous hydrodynamical and radiative transfer simulations, an embedded giant planet of several Jupiter masses is expected to launch significant spiral density waves in the disc, with a potentially observable arm outward from the planet location % if launched by a giant planet of a few Jupiter masses is present in the disc, based on previous hydrodynamical simulations and radiative transfer; 
\citep[e.g.][]{Zhu2015a,Dong2015a}. 
In particular, simulations of \citet{Zhu2015a} show that for discs with large inclinations the spiral is expected to be significantly brighter in full intensity than polarized intensity (see their Figure~17), which would be consistent with the absence of counterpart to Feature~\emph{iii} in polarized light.
Alternatively, Feature~\emph{iii} might be tracing thermal emission from shocks in the spiral rather than scattered light \citep[e.g.][]{Lyra2016}, which would also be consistent with the apparently large observed pitch angle ($\sim 27\pm4$\degr~after disc deprojection). %28deg
Yet another possibility is that Feature~\emph{iii} is a gap-crossing stream, which can be produced under certain companion mass ratio and orbital eccentricity conditions \citep[e.g.][]{Price2018}. % or spiral density waves \citep[][]{}
Either the spiral arm or the gap-crossing stream scenario would be compatible with an accreting protoplanet at the location of Feature~\emph{b} %, as suggested from tentative detection of H$\alpha$ emission 
\citep{Wagner2018a}.
Finally, we note that Feature~\emph{iii} roughly follows the inner edge of CO J=3-2 bulk emission in the South part of the disc, as seen with ALMA \citep{Long2018}.
However, new ALMA observations at higher angular resolution are required to confirm the possible connections between features seen in NIR and sub-mm wavelength images.

Altogether we cannot exclude the possibility that the extended features seen in our PCA-SADI image are authentic. 
New observations of PDS~70 making use of reference star differential imaging \citep[RDI; e.g.][]{Lafreniere2009,Ruane2017,Rameau2017} might confirm and better constrain these extended signals.
Whether a spiral arm or a gap-crossing stream is expected in the case of PDS~70 needs to be tested with dedicated hydrodynamical simulations, followed with radiative transfer predictions and post-processing. % (i.e.~forward modeling). 

\section{Summary} \label{Summary}

We presented the first application of combined spectral and angular differential imaging to VLT/SINFONI data.
We detailed the techniques we used to leverage both the spectral and angular diversity %to efficiently subtract speckle.
%We take advantage of the significant angular and spectral diversity present 
%in our data 
for an optimal stellar halo modeling and speckle subtraction using principal component analysis (PCA) in the case of a SINFONI dataset on PDS~70.
We then compared the results obtained when applying spectral differential imaging before angular differential imaging (PCA-SADI), or vice-versa (PCA-ASDI).
STIM maps \citep{Pairet2018} were used to identify significant features in our final images.
Our PCA-ASDI image reveals a point source at roughly the same location as the recently claimed protoplanet, while our PCA-SADI image shows both azimuthally extended features at a similar radius as the companion, and the faint back-scattered light from the far side of the disc. %that are reminiscent of those seen in the SPHERE images.  %in our final images, we pinpointed two flavours of our PCA-
Most features have counterparts in the SPHERE image, but a possible spiral arm or gap-crossing stream which appears connected to the location of the companion. % (Feature~\emph{iii}).

%Our images reveal the presence of a point-like source in the cavity and some tentative gap-crossing stream connecting the companion candidate and the outer disc.
%in agreement with the recent VLT/SPHERE detection.
%We also compare the behaviour of applying SDI before ADI (PCA-SADI) or vice-versa (PCA-ASDI) using synthetic spiral arms and fake companions injections. 
%We find that the former recovers better (azimuthally) extended features, while the latter is more sensitive to the detection of faint companions, even when surrounded by extended circumstellar signals.
%***This suggests that careful post-processing and similar quality final images as extreme-AO instrument SPHERE.

In order to better interpret the observed features and disentangle possible geometric biases inherent to the PCA-SADI and PCA-ASDI algorithms, we carried out a series of tests consisting in the post-processing of copies of our original datacube in which we injected either synthetic spirals (Section~\ref{SpiralInjectionTests}) or fake companions (Section~\ref{CompanionInjectionTests}).
%We carried out dedicated tests that probe the potential of PCA-SADI and PCA-ASDI to recover extended features and point sources, and 
We concluded that PCA-SADI recovers better azimuthally extended features, while PCA-ASDI is more sensitive to the detection of faint companions (or at least of companions with a similar spectrum as PDS~70~b), hence accounting for the differences in the PCA-SADI and PCA-ASDI images.
Therefore, we recommend the use of both PCA-SADI and PCA-ASDI for the post-processing of similar IFS datasets in the future. %, as they are more sensitive to different kind of features.
We also encourage the systematic use of dedicated tests (such as injection of synthetic extended features) when ADI is used.
%Forward modeling of a disc model is required to better assess the reliability of extended features identified in our image, in particular to confirm the possible gap-crossing stream or spiral arm. %connecting the edge of the outer disc and the protoplanet candidate.
%Proper forward modeling required -
%This is beyond the scope of this paper, and will be investigated in a forthcoming paper.

Based on our PCA-SADI/PCA-ASDI images and spiral injection tests alone, we cannot rule out the possibility that part of the signal from the location of the companion stems from a filtered extended structure. However, based on the independent ANDROMEDA detection at the location of PDS~70~b, we confirm that at least part of the signal comes from a point source.
Furthermore, both the spectrum and multi-epoch astrometry presented in \citetalias{Muller2018} argue in favor of an authentic protoplanet.
We hence adapted both the negative fake companion technique and our PCA-ASDI algorithm to infer reliable estimates of the astrometry,  and found the protoplanet to be at $r = 193.5 \pm 4.9$mas ($21.9 \pm 0.6$ au) and PA $= 158.7\degr \pm 3.0$\degr in our images, consistent with the astrometry reported in \citetalias{Keppler2018} and \citetalias{Muller2018} at other epochs. %{\bf ***Include ANDROMEDA result for astrometry???}.
We then extracted the spectro-photometry of the companion using two independent techniques: our negative fake companion algorithm and ANDROMEDA.
Both methods led to estimates of the contrast of the companion (with respect to the star) as a function of wavelength that are consistent with each other. However, the ANDROMEDA spectrum boasts both a higher spectral resolution and smaller uncertainties than the spectrum inferred by NEGFC (PCA-ASDI$_{\lambda}$), which we attribute to the ability of ANDROMEDA to ignore any bias from signals that are not point-like in nature. The NEGFC and ANDROMEDA spectra both agree with the SPHERE $K1$ photometric estimates obtained on 2016/05 and 2018/02, and the first epoch $K2$ measurement \citepalias{Keppler2018,Muller2018}. However, the second epoch $K2$ measurement is slightly fainter than our estimates. %and $> 7.1$mag in H band, hence with the SPHERE $K1$/$K2$ estimates). 
A detailed spectral analysis will be presented in a forthcoming paper (Christiaens et al. 2019, subm.~to ApJL).

Finally, we also adapted our PCA-SADI and PCA-ASDI algorithm to set constraints on the Br$\gamma$ line emission.
Assuming a similar Br$\gamma$ luminosity to accretion luminosity relationship as T-Tauri stars, our non-detection of Br$\gamma$ emission sets a limit on the mass accretion rate of the companion: $\dot{M_b} < 1.26 \times 10^{-7} \big[ \frac{5 M_{\rm Jup}}{M_b} \big] \big[ \frac{R_b}{R_{\rm Jup}}\big] M_{\rm Jup} $ yr$^{-1}$.
Considering both our non-detection of Br$\gamma$ and the tentative detection of H$\alpha$ emission, we infer that the visual extinction towards the companion should be smaller than $\sim$3.0 mag.

From a technical perspective, we used the full potential of an image-slicer based IFS not originally conceived to reach extremely high contrasts, and obtained final images of similar quality as more dedicated extreme AO-fed instruments such as VLT/SPHERE. %!!!
%AO-fed IFS like VLT/SINFONI or Keck/OSIRIS offer medium spectral resolution which is currently not available behind extreme AO systems neither on the ground nor in space.
%Our new images of the PDS~70 system suggest that optimal post-processing of IFS data obtained in pupil-stabilized mode can lead to similar quality final images as extreme-AO instruments such as VLT/SPHERE.
%***thanks to high spectral correlation of speckles
%{\bf The application of combined spectral and angular differential imaging on SINFONI leads to a larger gain in contrast than the application on lens-let based IFS \citep[see e.g.][]{Galicher2018}, which we interpret as due to the larger spectral correlation.}
Our results suggest that future IFS with a similar design but fed with extreme adaptive
optics systems or in space (e.g.~JWST/NIRSPEC) may reach even higher contrasts.

\section*{Acknowledgements}

VC is thankful to Sebastian Perez, Miriam Keppler and Roy van Boekel for constructive discussions. VC also acknowledges Zachary Long and Mike Sitko for sharing their SpeX spectrum of PDS~70, %Andre M\"uller for the SPHERE spectrum of PDS~70~b, 
and Alexander Heger for the use of Oxygen.
VC and SC acknowledge support from the Millennium Science 
Initiative (Chilean Ministry of Economy) through grant RC130007. 
%The research leading to these results has received 
VC and OA acknowledge funding from the European Research Council under the European Union's Seventh Framework Programme (ERC Grant Agreement No.~337569) and from the French Community of Belgium through an ARC grant for Concerted Research Action.
%OA acknowledges funding from F.R.S.-FNRS.
We also acknowledge funding from the Australian Research Council via DP180104235.
This work was supported by the Multi-modal Australian ScienceS Imaging and Visualisation Environment (MASSIVE) (\url{www.massive.org.au}). 

% VC also acknowledges Miriam Keppler and Roy van Boekel for sharing their paper.
% VC also acknowledges support from CONICYT through CONICYT-PCHA/Doctorado Nacional/2016-21161112

%%%%%%%%%%%%%%%%%%%%%%%%%%%%%%%%%%%%%%%%%%%%%%%%%%



\bibliographystyle{mnras}
\bibliography{PDS70} % if your bibtex file is called example.bib

%\begin{thebibliography}{99}
%\end{thebibliography}

%%%%%%%%%%%%%%%%%%%%%%%%%%%%%%%%%%%%%%%%%%%%%%%%%%

%%%%%%%%%%%%%%%%% APPENDICES %%%%%%%%%%%%%%%%%%%%%

\appendix

\section{SNR maps versus STIM maps} \label{SNR_vs_STIM}

\begin{figure*}
	\centering
	\includegraphics[width=0.99\textwidth]{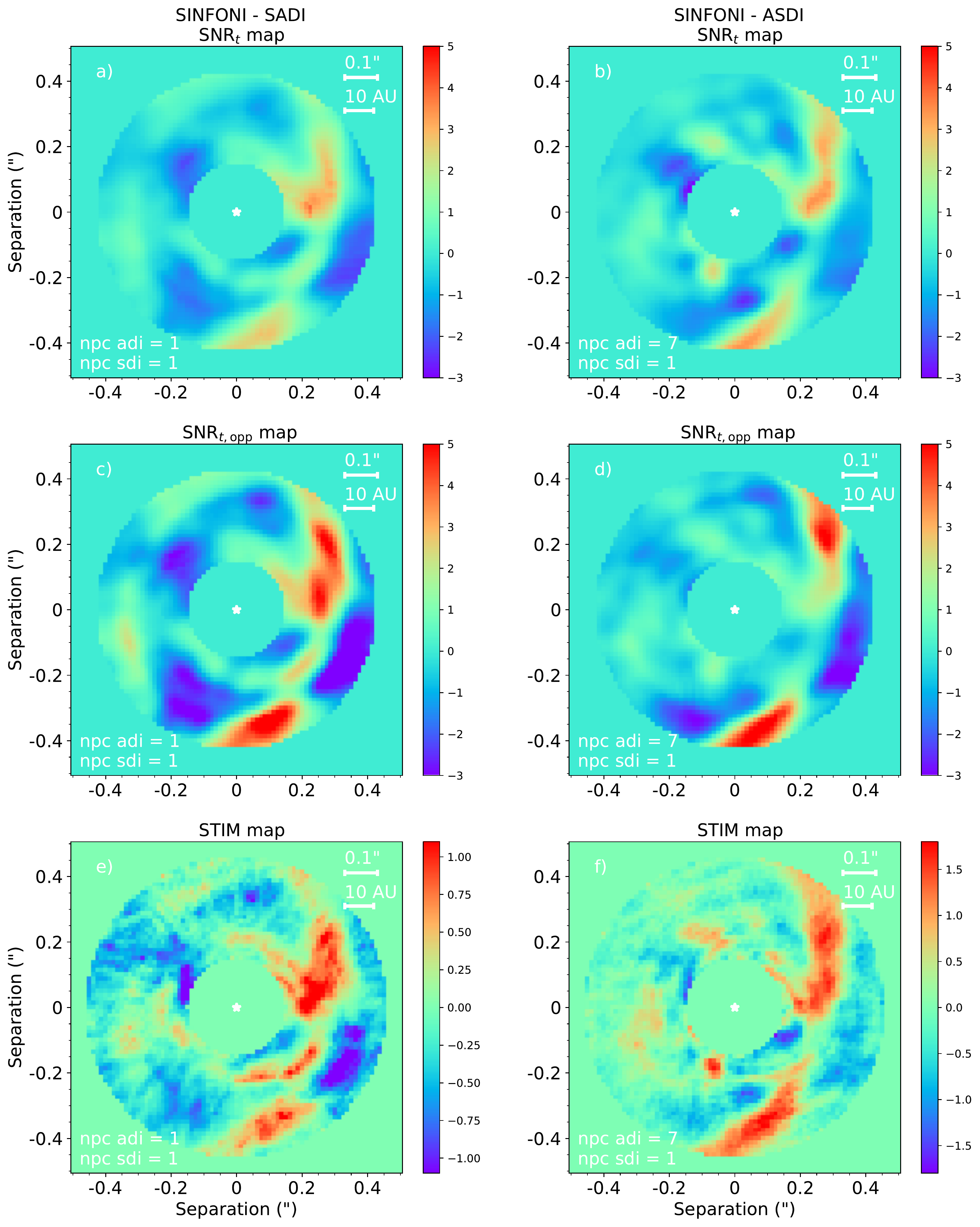}
    \caption{Comparison between the SNR$_t$ maps \citep[top row;][]{Mawet2014}, SNR$_{t,\mathrm{opp}}$ maps (middle row) and STIM maps \citep[bottom row;][]{Pairet2018} for the best PCA-SADI and PCA-ASDI reductions. SNR$_{t,\mathrm{opp}}$ maps are similar to SNR$_t$ maps, except that the independent apertures used to estimate the noise at each radial separation are considered in the image obtained using opposite derotation angles. The latter can be considered a proxy of the residual speckle noise since no authentic signal adds up constructively while the time dependence of residual speckle is preserved. In contrast to SNR maps, STIM maps are not limited by the lack of independent apertures at short separation for the estimation of the local residual speckle noise.}
    \label{SNR_STIM_maps}
\end{figure*}

Despite a widespread use of advanced differential imaging techniques in the HCI community, assessing the significance of signals found in post-processed images remains challenging and depends on assumptions regarding the statistics of residual speckle intensities \citep[e.g.][]{MaroisLafreniere2008,Mawet2014,JensenClem2018}. % \citet{Mawet2014}.
\citet{Mawet2014} suggested to compute the signal-to-noise ratio (SNR) at each pixel as the ratio between the flux measured in a FWHM-wide aperture at the pixel location and the standard deviation of the fluxes measured in independent FWHM-wide apertures at the same radial separation, corrected by a radius-dependent factor reflecting small sample statistics (Student $t$ statistics).
SNR$_t$ maps computed in such a way are now routinely used.
However, they assume the presence of at most one point source at each radial separation from the star.
In the case of PDS~70, the bright, well-characterized, edge of the outer disc extends over a continuous range of projected radii from the NW to the S of the star \citep[][\citetalias{Keppler2018,Muller2018}]{Hashimoto2012}.
Furthermore, some additional extended features might also stem from within the large annular gap \citepalias{Muller2018}.
These bright extended features affect the flux measurements in the independent apertures, and hence artificially increase the noise level estimated at each radius.
As a consequence, according to the classical SNR$_t$ maps corresponding to Figure~\ref{PCA-ASDI_final_img}a and b (Figure~\ref{SNR_STIM_maps}a and b, resp.), no significant signal (SNR$>$ 5) is present in our images, despite the presence of visually conspicuous signals coinciding with features previously identified in different datasets.

Therefore, we adapted the classical SNR$_t$ definition in order to estimate a more appropriate SNR map in the presence of extended disc features in the image. 
Instead of estimating the noise at each radius using independent apertures in the final image, we consider the same apertures but in the image obtained using the \emph{opposite} values of parallactic angles for derotation of the residual frames. %We proceeded in two steps: we first estimated a noise map associated to each image, and then used it for measuring the standard deviation of the fluxes in independent apertures at each radial separation.
%The noise maps were produced by post-processing the data with each algorithm using the \emph{opposite} values of parallactic angles during derotation of the stellar-halo subtracted frames, i.e.~before final stacking.
The images produced in such a way preserve the time dependence of residual speckle, while not constructively adding authentic circumstellar signal (point-like or extended), and are hence good proxies of the residual speckle noise level in the image \citep[e.g.][]{MaroisLafreniere2008,Wahhaj2013}.
Contrary to the classical SNR$_t$ maps, these SNR$_{t,\mathrm{opp}}$ maps now suggest that parts of the forward-scattered edge of the outer disc are significant (Figure~\ref{SNR_STIM_maps}c and d).
Nonetheless, this procedure still suffers from the small number of independent apertures at short radial separation.

An alternative way to assess the significance of signals in HCI post-processed images is to use standardized trajectory intensity mean (STIM) maps \citep{Pairet2018}.
STIM maps leverage the temporal variation of pixel intensities resulting from the trajectory of speckles in the derotated cube of residual images (i.e.~the cube of images obtained after PCA modeling and subtraction, and subsequent alignment with North up and East left).
These 2D detection maps are defined as $\mu (x_{i,j})/\sigma (x_{i,j})$, where $\mu (x_{i,j})$ and $\sigma (x_{i,j})$ are the mean and standard deviation of trajectory $x_{i,j}$ (i.e.~the transversal slice at location ($i$,$j$)) throughout the derotated cube of residual images, respectively.
Pixels containing authentic circumstellar signals (disc or planet) present a significantly higher value of $\mu(x_{i,j})/\sigma(x_{i,j})$ than pixels with no circumstellar flux contribution, due to (1) the relatively more constant (and higher) intensity of pixels containing circumstellar signal, and (2) the relatively higher standard deviation for pixels devoid of circumstellar signal due to the trajectory of speckles crossing these pixels in the derotated cube.
By construction, STIM maps address the two main drawbacks of applying SNR$_t$ maps to our PDS~70 images: the presence of bright circumstellar signals spanning a range of radii and small number of independent apertures at short separation. 
The original definition of STIM maps was made in the context of 3D ADI datacubes \citep{Pairet2018}.
However, it is conceptually similar to apply it to SDI data; the trajectories are only radial instead of azimuthal.
%Furthermore, PCA-SADI and PCA-ASDI consist in two consecutive PCA modeling and subtraction, which lead to a better whitening of the noise in the cube of residual images.
Our STIM maps for PCA-SADI and PCA-ASDI are thus computed on the (3D) cube of residual images obtained after the second PCA (ADI or SDI, resp.), for which the only change compared to the original STIM map definition is that it benefitted from a better whitening of the noise than using only a single PCA-ADI or PCA-SDI.
They are shown in Figures \ref{PCA-ASDI_final_img}c and d, and \ref{SNR_STIM_maps}e and f.

\begin{figure}
	\centering
	\includegraphics[width=0.49\textwidth]{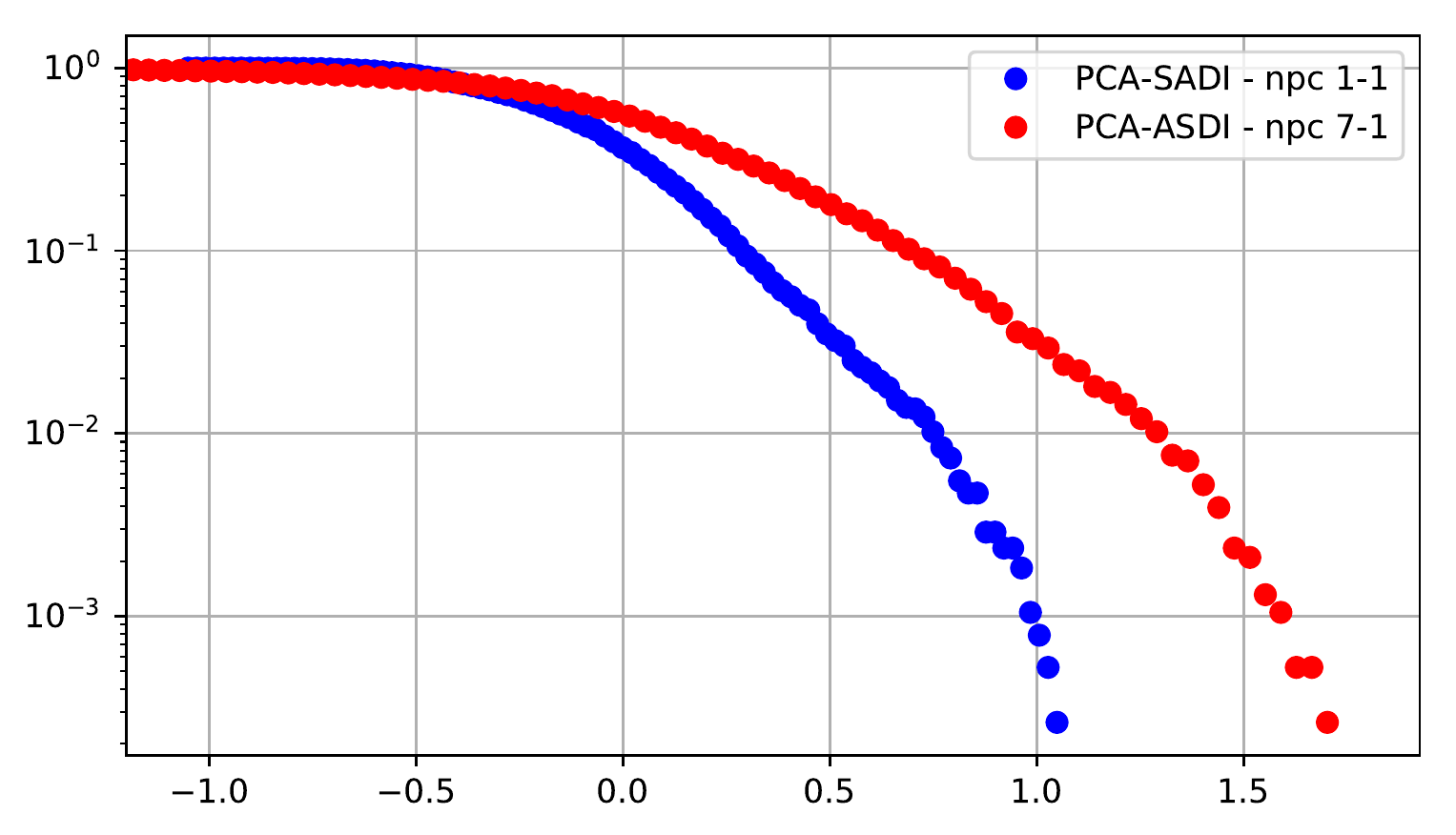}
    \caption{Cumulative fraction of pixel values in the inverse STIM maps obtained with PCA-SADI ($n_{\rm pc}^{\rm ADI}  = 1$; $n_{\rm pc}^{\rm SDI}  = 1$) and PCA-ASDI  ($n_{\rm pc}^{\rm ADI}  = 7$; $n_{\rm pc}^{\rm SDI}  = 1$).
    The plot suggests that 1.1 and 1.8 are appropriate thresholds for the STIM map color scales shown in Figure~\ref{PCA-ASDI_final_img}c and d, respectively.}
    \label{Inv_STIM_maps}
\end{figure}

In order to interpret the STIM maps and identify significant features, it is necessary to compute a STIM map using opposite parallactic angles (hereafter referred to as \emph{inverse STIM map}).
As explained for the SNR$_{t,\mathrm{opp}}$ map, the inverse STIM map is expected to be a good proxy of the residual speckle noise in the (correctly derotated) STIM map.
Figure \ref{Inv_STIM_maps} shows the cumulative fraction of pixel values in the inverse STIM maps for PCA-SADI ($n_{\rm pc}^{\rm SDI}  = 1$; $n_{\rm pc}^{\rm ADI}  = 1$) and PCA-ASDI ($n_{\rm pc}^{\rm ADI}  = 7$; $n_{\rm pc}^{\rm SDI}  = 1$). 
\citet{Pairet2018} suggested to use this kind of plot to identify the threshold pixel value above which signal is very unlikely to stem from the random combination of residual speckle in the (correctly derotated) STIM maps. 
In view of the cumulative fraction distributions, we consider threshold values of 1.1 and 1.8 for the PCA-SADI ($n_{\rm pc}^{\rm SDI}  = 1$; $n_{\rm pc}^{\rm ADI}  = 1$) and PCA-ASDI ($n_{\rm pc}^{\rm ADI}  = 7$; $n_{\rm pc}^{\rm SDI}  = 1$) STIM maps, resp. (Figure~\ref{PCA-ASDI_final_img}c and d, resp.).

\section{Reliability of spectro-astrometry inferred by NEGFC} \label{tests_spectrophotometry}

\citet{Wertz2017} proposed to estimate the astrometric errors associated to residual speckle noise by injecting a statistical number of fake companions at the same radial separation and contrast as the companion but at different PAs, and measuring the deviations between parameters used for injection and NEGFC-inferred parameters. 
We followed the same method and injected 42 fake companions in separate copies of our original cube where the protoplanet was subtracted using the best estimates of $r$, PA and contrast($\lambda$) inferred in Section~\ref{NEGFCresults}. %***CHANGE FOR 42 ***
Given the bright forward scattered edge of the disc to the W and SW of the image, we only injected the fake companions between PA=-30\degr~and PA=170\degr, at 10\degr~intervals, at the same radial separation as the protoplanet. %***CHANGE FOR 42 ***
%However, this was not appropriate to the case of PDS 70, due to the presence of strong circumstellar signals at similar separation as PDS~70~b (such as Features~\emph{i},~\emph{ii} and~\emph{iii} identified in Figure~\ref{PCA-ASDI_final_img}).
%Given the tests presented in Appendix~B of \citet{Christiaens2018} and similarity between the HD~142527 and PDS~70 datasets, we adopted a similar strategy as in that
%Therefore, the residual speckle noise uncertainty was rather estimated in a similar way as described in Appendix~B of \citet{Christiaens2018}.
We carried out the NEGFC optimization for 8 values of $n_{\rm pc}^{\rm ADI}$ (from 3 to 10 included), for which the companion was visually conspicuous in the final image (Figure~\ref{npcADI_evo}). %, two slightly different aperture sizes (0.5 and 0.6 FWHM), and two figures of merit (minimization of absolute sum and standard deviation of residual pixel values in the aperture at the companion location), hence 32 runs in total.
%***Improvement with VAR-collapse?***
We then found the optimal $n_{\rm pc}^{\rm ADI}$ that minimized the mean deviation between the true ($r$, PA) used for the injection and the NEGFC estimates for the 42 fake companions. %***CHANGE FOR 42 ***
Then, we considered the standard deviation of these 42 deviations as the astrometric uncertainty associated to residual speckle noise.  %***CHANGE FOR 42 ***
We found uncertainties of 4.8~mas and 2.9\degr, for $r$ and PA respectively.  %***CHANGE FOR 42 ***
%The uncertainties correspond to the standard deviation of $r$ and PA estimates throughout those 32 runs.
%These residual speckle noise uncertainties dominate the error budget (3.7mas and ). 

\begin{figure}
	\centering
	\includegraphics[width=0.49\textwidth]{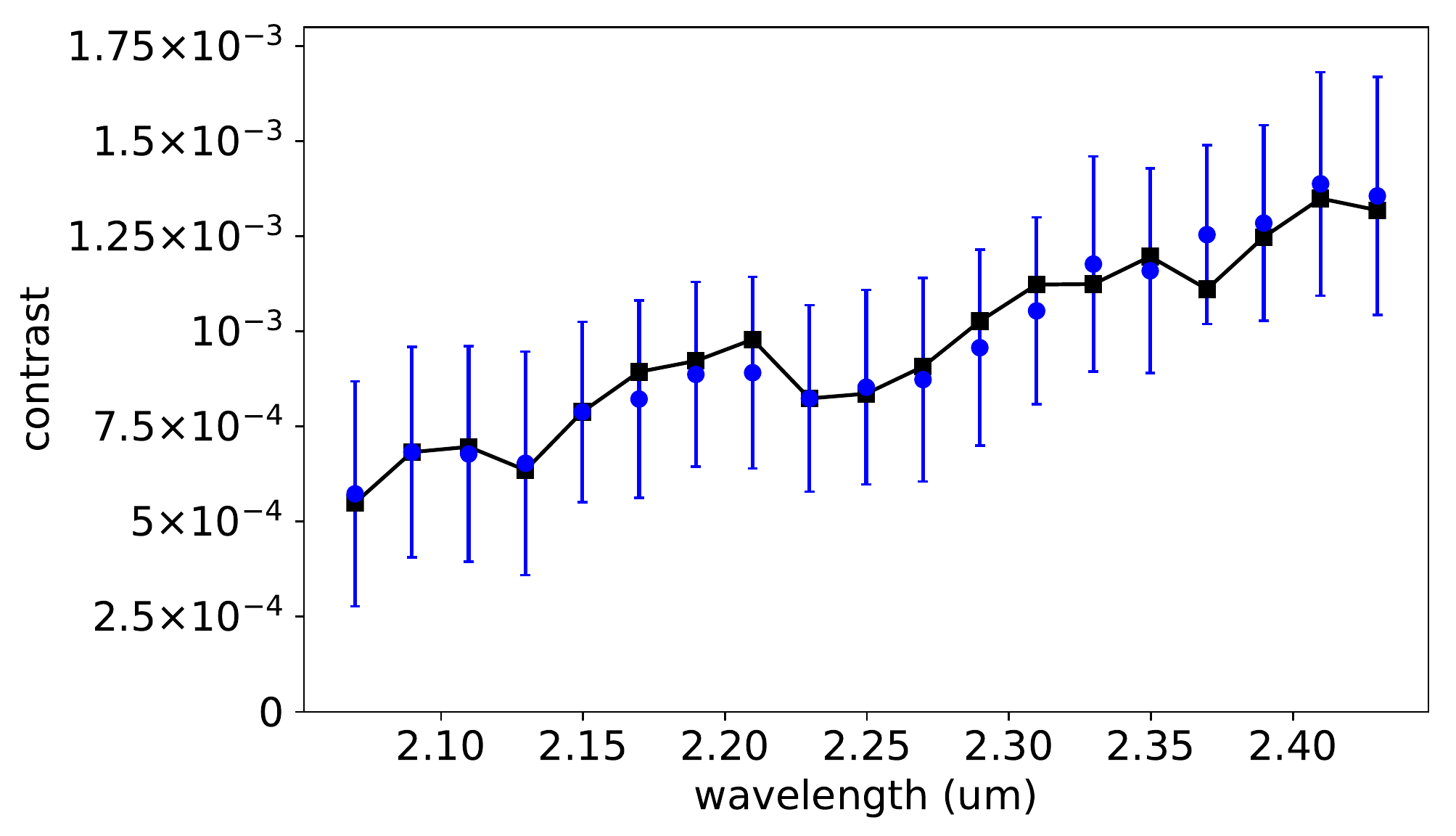}
    \caption{\emph{Black squares}: spectrum used for the injection of fake companions (same as inferred for protoplanet PDS 70 b). \emph{Blue circles}: Median flux inferred by NEGFC (PCA-ASDI$_{\lambda}$) over the 42 injected fake companions, for the number of principal components that minimize the deviations. Error bars represent the standard deviation of the deviations between the injected and estimated contrast($\lambda$) with the optimal $n_{\rm pc}^{\rm ADI}$.} %***CHANGE FOR 42 ***
    \label{Injected_vs_retrieved_spectrum}
\end{figure}

To ensure that the spectrum inferred for the companion is not distorted by the PCA-ASDI$_{\lambda}$ algorithm, 
we followed a similar procedure as above for the inferred contrast. 
%We injected {\bf 42} fake companions at the same contrast as a function of wavelength as inferred for the protoplanet.
Given the prior lack of knowledge of the optimal value of $n_{\rm pc}^{\rm ADI}$, we injected 42 fake companions at the median contrast as a function of wavelength inferred by NEGFC (PCA-ASDI$_{\lambda}$) for the authentic companion for values of $n_{\rm pc}^{\rm ADI}$ ranging in $[3, 10]$.  %***CHANGE FOR 42 ***
Then, we applied NEGFC (PCA-ASDI$_{\lambda}$) to estimate the spectrum of each of the fake companions.
We considered either a median collapse or a variance-weighted average of residual images, and noticed a slight decrease of deviations between recovered and injected spectra using the variance-weighted average.
For each binned spectral channel, we found the optimal $n_{\rm pc}^{\rm ADI}$ that minimized the median deviation between the true injected flux and the NEGFC estimates of the flux for the 42 fake companions. %***CHANGE FOR 42 ***
This optimal $n_{\rm pc}^{\rm ADI}$ was then used to infer the final contrast as a function of wavelength of the companion (Figure~\ref{FinalContrast}).
The error bars in Figure~\ref{Injected_vs_retrieved_spectrum} are the standard deviation of the deviations between the injected and estimated contrast($\lambda$) using the optimal $n_{\rm pc}^{\rm ADI}$ for each channel, and are also used in Figure~\ref{FinalContrast}.
Figure~\ref{Injected_vs_retrieved_spectrum} shows that the spectrum of fake companions inferred by NEGFC (PCA-ASDI$_{\lambda}$) is consistent with the true injected spectrum, for all binned spectral channels.

%In particular, the spectrum is not significantly distorted compared to the injected one, we do notice that the inferred spectrum is slightly under-estimated for a large
%which casts confidence on the shape of the spectrum extracted for the protoplanet candidate.
% If a similar bias affects the spectrum inferred for the companion, it would mean its spectrum is even redder than inferred, hence further arguing for the presence of a circumplanetary disc.

\section{The ASDI+Br$\gamma$ and SADI+Br$\gamma$ algorithms} \label{AppBrG}

\emph{ASDI+Br}$\gamma$ is similar to what is referred to as \emph{ASDI} in \citet{Close2014} and \emph{SDI+} in \citet{Wagner2018a}, with the difference that we correct for algorithmic flux losses in each channel by measuring the throughput with injected fake companions. Our detailed procedure is as follows:
\begin{enumerate}
\item We first extracted a raw spectrum of the star using 1-FWHM apertures in the median $H$+$K$ spectral cube before ADI processing. No stellar emission line standing out from the continuum was found around the Br$\gamma$ line, but we found an absorption line slightly shortward (at $\sim 2.161 \mu$m).
\item We defined our \emph{line channels} as all spectral channels lying within 4 spectral FWHM from the center of the Br$\gamma$ line, where the assumed gaussian spectral FWHM of the line was varied between 7.5, 15 and 30 \AA~(i.e.~we considered 6, 12 and 24 spectral channels). 
An identical number of \emph{continuum channels} were considered directly longward from the last line channel (to avoid the stellar absorption line at shorter wavelength).
\item The continuum channels were rescaled both physically and in flux to match the size and flux of a weighted average of the line channels. The weights followed a gaussian centered on the Br$\gamma$ line, with the 3 tested values of line FWHM, and were normalized to 1.
\item We applied PCA-ADI in concentric annuli, on both the line and continuum channels, individually. 
\item In each channel, we estimated the throughput of the algorithm as a function of radial separation. This was obtained at each radial separation with the injection of 7 fake companions at different azimuths, considering the throughput as the median throughput value out of those 7 flux ratio measurements (the algorithmic throughput is the ratio of the retrieved flux to the injected flux).
\item We divided the PCA-ADI frames by the throughput map computed in the previous step.
\item The Br$\gamma$ image, $I_{\textrm{Br}\gamma}$, was calculated as a weighted average of the PCA-ADI frames of the line channels (Figure~\ref{BrGImages}a). The weights were the same as in step (iii). % follow a gaussian with a FWHM of 7.5 \AA~(as the Br$\gamma$ emission of HD 142527; Mendigutia 2014) centered on the peak wavelength of Br$\gamma$ (2.1661 $\mu$m), and are such that their sum equals 1. 
%In practice, only ***SINFONI spectral channels are considered for the weighted-average (channels within $\sim$4 FWHM from the center of the line). We considered 3 different line FWHM values: 7.5, 15 and 30 $\AA$.
The continuum image, $I_{\textrm{cont}}$, was computed as the median of the PCA-ADI frames of the continuum channels  (Figure~\ref{BrGImages}b).
\item We obtained the \emph{ASDI+Br}$\gamma$ image (Figure~\ref{BrGImages}c) by subtracting the continuum image from the Br$\gamma$ image: $I_{\textrm{ASDI+Br}\gamma} = I_{\textrm{Br}\gamma} - I_{\textrm{cont}}$.
\item Steps (i) to (viii) were repeated varying the number of principal components ($n_{\rm pc}^{\rm ADI} $) between 1 and 10. A lower $n_{\rm pc}^{\rm ADI} $ led to more correlated residual speckle noise between spectral channels, but a larger $n_{\rm pc}^{\rm ADI} $ produced a better raw contrast in each PCA-ADI image. We determined the value of $n_{\rm pc}^{\rm ADI} $ that optimized sensitivity for Br$\gamma$ line emission at the location of the companion to be 7, as this value minimized the throughput-corrected noise at the radial separation of the companion in $I_{\textrm{ASDI+Br}\gamma}$.
\end{enumerate}

By contrast, \emph{SADI+Br}$\gamma$ is similar to what is referred to as \emph{ASDI} in \citet{Rameau2015}. Our detailed procedure was as follows: 
\begin{enumerate}
\item We applied steps (i) to (iii) defined for the \emph{ASDI+Br}$\gamma$ algorithm. 
\item We calculated the pre-ADI Br$\gamma$ image as a weighted average of the line spectral channels. %(with weights determined in step 3). 
Similarly, the pre-ADI continuum image is computed as the median of the continuum spectral channels. %(identified in step 2). 
\item For each spectral cube of our observing sequence, the pre-ADI continuum image was subtracted from the pre-ADI Br$\gamma$ image. This led to a 3D (ADI) cube consisting of continuum subtracted Br$\gamma$ images.
\item PCA-ADI was applied to that cube in concentric annuli. We also applied PCA-ADI on the continuum and (non-continuum subtracted) Br$\gamma$ cubes.
\item We estimated the throughput of PCA-ADI as in step v of the \emph{ASDI+Br}$\gamma$ algorithm.
\item We divided the PCA-ADI frames by the throughput map computed in the previous step. This led to a final \emph{SADI+Br}$\gamma$ image (Figure~\ref{BrGImages}f), and to both continuum and (non-continuum subtracted) Br$\gamma$ PCA-ADI images ($I_{\textrm{cont}}$ and $I_{\textrm{Br}\gamma}$, Figure~\ref{BrGImages}d and e, resp.), all of which were throughput-corrected.
\item %This step is identical to step ix of \emph{ASDI+Br}$\gamma$: 
We repeated the process (steps i to vi) for different $n_{\rm pc}^{\rm ADI}$ until we found the value of $n_{\rm pc}^{\rm ADI}$ that minimizes the throughput-corrected noise. The optimal $n_{\rm pc}^{\rm ADI} $ for \emph{SADI+Br}$\gamma$ was 2.
\end{enumerate}

%%%%%%%%%%%%%%%%%%%%%%%%%%%%%%%%%%%%%%%%%%%%%%%%%%

% Don't change these lines
\bsp	% typesetting comment
\label{lastpage}
\end{document}